\documentclass[sigconf]{acmart}

\AtBeginDocument{%
  \providecommand\BibTeX{{%
    \normalfont B\kern-0.5em{\scshape i\kern-0.25em b}\kern-0.8em\TeX}}}

\copyrightyear{2022} 
\acmYear{2022} 
\setcopyright{acmlicensed}\acmConference[MuC '22]{Mensch und Computer 2022}{September 4--7, 2022}{Darmstadt, Germany}
\acmBooktitle{Mensch und Computer 2022 (MuC '22), September 4--7, 2022, Darmstadt, Germany}
\acmPrice{15.00}
\acmDOI{10.1145/3543758.3543947}
\acmISBN{978-1-4503-9690-5/22/09}

\usepackage{multirow}
\usepackage{makecell}
\usepackage{tabularx}

\usepackage{caption}
\usepackage{subcaption}

\usepackage[capitalize]{cleveref}

\definecolor{DanielsColor}{rgb}{0.9,0.6,0.1}

\definecolor{NiklasColor}{rgb}{0.9,0.3,0.9}

\definecolor{FloriansColor}{rgb}{0,0.3,0.9}

\newcommand{\stdiabbr}[0]{\textit{Base}}
\newcommand{\stdiabbrs}[0]{\stdiabbr\space}

\newcommand{\cgt}[0]{\textit{Continuous Generated Text}}
\newcommand{\cgts}[0]{\cgt\space}
\newcommand{\cgtabbr}[0]{\textit{Cont}}
\newcommand{\cgtabbrs}[0]{\cgtabbr\space}

\newcommand{\wws}[0]{\textit{Writing with Suggestions}}
\newcommand{\wwss}[0]{\wws\space}
\newcommand{\wwsabbr}[0]{\textit{Sugg}}
\newcommand{\wwsabbrs}[0]{\wwsabbr\space}

\newcommand{\p}[1]{$p #1$}

\newcommand{\glmmci}[5]{$\beta$=#1, SE=#2, CI$_{95\%}$=[#3, #4], \p{#5}}

\newcommand{\ztest}[4]{#1 vs. #2: Z = #3, $p #4$}

\newcommand{\fmann}[3]{$\chi^2$ (2, N=#1) = #2, \p{#3}}
\newcommand{\wilcox}[2]{Z = #1, \p{#2}}

\newcommand{\meandesc}[1]{$M_#1$}

\newcommand{\lastaccessed}{\textit{last accessed 10.03.2022}}

\begin{document}


\title{Suggestion Lists vs. Continuous Generation: Interaction Design for Writing with Generative Models on Mobile Devices Affect Text Length, Wording and Perceived Authorship}

\renewcommand{\shorttitle}{Suggestion Lists vs. Continuous Generation: Interaction Design for Writing with Generative Models}


\author{Florian Lehmann}
\orcid{0000-0003-0201-867X}
\authornote{Both authors contributed equally to this research.}
\email{florian.lehmann@uni-bayreuth.de}

\author{Niklas Markert}
\email{niklas.markert@uni-bayreuth.de}
\authornotemark[1]
\affiliation{%
  \institution{Department of Computer Science, University of Bayreuth}
  \city{Bayreuth}
  \country{Germany}
}

\author{Hai Dang}
\orcid{0000-0003-3617-5657}
\email{hai.dang@uni-bayreuth.de}
\affiliation{%
  \institution{Department of Computer Science, University of Bayreuth}
  \city{Bayreuth}
  \country{Germany}
}

\author{Daniel Buschek}
\orcid{0000-0002-0013-715X}
\email{daniel.buschek@uni-bayreuth.de}
\affiliation{%
  \institution{Department of Computer Science, University of Bayreuth}
  \city{Bayreuth}
  \country{Germany}
}

\renewcommand{\shortauthors}{Lehmann and Markert, et al.}

\begin{abstract}
Neural language models have the potential to support human writing. However, questions remain on their integration and influence on writing and output. To address this, we designed and compared two user interfaces for writing with AI on mobile devices, which manipulate levels of initiative and control: 1) Writing with continuously generated text, the AI adds text word-by-word and user steers. 2) Writing with suggestions, the AI suggests phrases and user selects from a list. In a supervised online study (N=18), participants used these prototypes and a baseline without AI. We collected touch interactions, ratings on inspiration and authorship, and interview data. With AI suggestions, people wrote less actively, yet felt they were the author. Continuously generated text reduced this perceived authorship, yet increased editing behavior. In both designs, AI increased text length and was perceived to influence wording. Our findings add new empirical evidence on the impact of UI design decisions on user experience and output with co-creative systems. 
\end{abstract}



\maketitle
\section{Introduction}

Text input with touch interfaces, such as mobile devices like smartphones is cumbersome compared to text input on laptop or desktop computers \cite{palin19, varcholik12}. A basic approach to support the user with touch text input is giving suggestions based on language models \cite{goodman02}. This kind of support ships with the default keyboard software on every smartphone. Third-party keyboards such as \textit{SwiftKey}\footnote{\url{https://www.microsoft.com/en-us/swiftkey} \lastaccessed} or \textit{Fleksy}\footnote{\url{https://www.app.fleksy.com} \lastaccessed} provide even more support. Such keyboards do not only give word suggestions while typing, but have also auto-correction implemented. The main goal of such approaches is to reduce effort of the user by minimizing touch input. The underlying methods and technologies of these features stem from the domain of natural language processing (NLP). One approach for building these models are neural networks. A special form of these neural networks are Transformer networks \cite{vaswani17} and can be trained, for example, for question answering, translation, or text generation. From a human-computer perspective, we see a particular potential of models that generate text to influence the way we write. 

These generative models could be applied for writing text on mobile devices to further reduce the input effort. However, it remains unclear how to best design user interfaces that empower people to interact with text generation and how such design choices affect the text output as well as the experienced human-AI relationship. This motivates our central research question of this paper: \textit{How do fundamental human-AI roles, as embedded in UI and interaction design, influence text entry behavior, experience, and output?}

To make this measurable, we follow a comparative design and evaluation approach: In particular, we introduce a novel interaction method designed to reverse the ``roles'' of human and AI writers in the current state of the art (and industry): In this design, the AI continuously adds text word-by-word ``live'' and puts the user in the role of steering and editing the generation, rather than writing. In a user study, we compare this to both a non-AI baseline with manual text entry and to the common design of AI text as a suggestion list. This comparative experience for users allows us to address questions regarding perceived authorship, inspiration, efficiency, interaction behavior, and the overall human-AI relationship. 

Concretely, we created two prototypes that allow us to compare 1) writing with suggestions of sentence completions and 2) continuously generated text. In both prototypes, the text was generated with the neural language model GPT-2. We conducted a supervised, within-subject online study with 18 participants. They wrote a birthday greeting and a text about a vacation with each of these prototypes. To record a baseline, we also let them write text manually. We measured interactions, collected ratings, and conducted interviews on various aspects of the users interacting with the generative models with our input methods.

Our findings reveal aspects that are important when designing and implementing for interactions with generative text models regarding efficiency, inspiration, and editing behavior. In our comparison, we found that writing with AI, that continuously writes text live, leads to high costs for correcting semantically imprecise generations that did not fit expectations. Despite this higher activity, people perceived themselves as authors less since they edited instead of drafting. In contrast, writing with selecting from a set of AI generated phrase suggestions leads to reduced keyboard input yet higher perceived authorship. In both methods, users perceived an influence of the generated text on the wording.
We discuss these results considering the level of control, initiative, and roles.

With this work, we contribute: 1) Two prototypes for writing with AI on mobile devices. In particular, we introduce a novel interaction where the user steers the written text instead of actively writing it.  2) A study comparing these prototypes with an extensive analysis of interaction data as well as subjective ratings and feedback. 3) We give design recommendations based on our findings, for example, that a change in role can be used to make the user believe to be the author of a document, and outline implications for future research. 

Alongside with this paper, we make the data and study material publicly available to support future research on human interaction with NLP text models. \footnote{Data and study material is publicly available on OSF: \url{https://osf.io/p976a/}}

\section{Related Work}\label{sec:related_work}

\subsection{Writing with Language Models on Mobile Devices}
Making writing on mobile devices more efficient and comfortable is a challenging and important topic in HCI \cite{kristensson14}.

A popular approach to address this challenge is to rely on methods from NLP, for example language models of varying complexities. Simple (letter) frequency models can be used, for example, to build adaptive keyboards to reduce input errors \cite{alfaraj09, gunawardana10}. Recently, language models have also been employed for improving error correction, by automatically predicting where to place an entered correction text \cite{cui20, zhang19}.

The most widely known application of language models for mobile keyboards are word or phrase suggestions and auto-correction. These techniques are implemented in almost all major soft-keyboards, for example in Apple's standard iOS keyboard\footnote{\url{https://support.apple.com/guide/iphone/type-with-the-onscreen-keyboard-iph3c50f96e/ios} \lastaccessed}, Google's \textit{Gboard}\footnote{\url{https://play.google.com/store/apps/details?id=com.google.android.inputmethod.latin&hl=en&gl=US} \lastaccessed} or Microsoft's \textit{SwiftKey}.

In particular, the use of \textit{text suggestions} is interesting for our work: Suggestions provide the chance to increase efficiency by enabling users to select a word or phrase instead of typing it. In addition, selecting a suggestion eliminates the possibility of typos.
There are several studies on the use of suggestions. Some are more general, not focused on mobile devices, such as a study about the number of suggestions provided during email writing \cite{buschek21}.
Others are focused on mobile devices, such as studies on the influence of emotional state on selecting suggestions \cite{ghosh19} or suggestion presentation \cite{quinn16}. Other work examined suggesting phrases, in contrast to words \cite{arnold16}.

Language models are also used to suggest complete (short) replies: For example, \textit{Smart Reply}~\cite{kannan16} analyses an incoming email and suggests three short responses that can be selected and sent without any manual typing. 

This wide range of research on language models in (mobile) text entry motivates our investigation here. However, in contrast to the focus on efficiency that is currently predominant in the related work, we focus on effects of interaction design on user experience and output when writing ``together'' with AI (neural language models). To do this, we make design choices that are not motivated by efficiency, but rather by exploring different human-AI relationships. Ultimately, we envision that these two lines of investigation can be combined, in order to design efficient mixed-initiative \cite{horvitz99} text input systems with more explicitly considered user experience and output properties.
 
\subsection{Text Generation with Language Models}

Language Models model large text corpora as probability distributions \cite{bengio08}. They are typically trained to predict the next word in a text sequence. 
Two well-known representatives of such models, that also appear in recent HCI work~\cite{buschek21, lee_coauthor_2022, singh22}, are GPT-2\footnote{\url{https://huggingface.co/gpt2} \lastaccessed} \cite{radford19} and its successor GPT-3 \cite{brown20}. 

One way to use language models for creating text is by showing their output to the user as suggestions. These suggestions mostly contain single words or longer text phrases. 
The work of \citet{chang21} shows that it is also possible to give users ``topics'' to select instead of directly choosing the full suggestions. 
This could be especially useful on mobile devices, as there is only limited space to display longer suggestion texts.

A related practical example is Google's \textit{Smart Compose} which displays automatic inline suggestions (one at a time) to complete the current phrase or sentence while writing an email \cite{chen19}.

It is interesting to compare \textit{Smart Compose} with \textit{Smart Reply} (see above) since they embed different roles for user and AI: In \textit{Smart Compose}, the user is the main writer and the AI acts in an assisting role. In contrast, in \textit{Smart Reply}, the AI writes the different short answers for the incoming mail and the user only has to choose (and possibly edit) the preferred one. 

However, in most of the literature on human-AI writing, the role distribution keeps the user as the main writer and the AI as some sort of helper and/or corrector. Our work in this paper challenges this with an alternative design, in which the AI writes ``freely'' and the user corrects and edits. Our view here is not that this is a more practical alternative for everyday use. We rather see it as a point of contrast which enables us to explore, quantify and evaluate how human-AI roles can be explicitly manipulated through interaction design and how this affects experience and output.

Nevertheless, our alternative design here is not far-fetched and practically relevant, as recent industry examples show: For instance, a related product is \textit{Flowrite}\footnote{\url{https://www.flowrite.com} \lastaccessed}: Users select a theme and provide keywords for the content of an email, which a language model then turns into a full draft. One difference between this product and our studied design here is that we add the generated text word-by-word (and not at once) to emphasize the impression of the system as ``writing''. Moreover, this gives users the opportunity to react mid-text (e.g. interrupt generation and edit).

\subsection{Impact of Writing with AI}
Writing with the help of language models can also impact \textit{what} is written, that is, the output text.

For example, the aforementioned AI text suggestions may not only improve text input, but can also provide incentives for exploring new wording or topics. There are several papers about the use of suggestions in creative writing \cite{calderwood20, roemmele15}. The work by \citet{arnold20} shows that typing with suggestions impact the writing and results in a shorter and more predictable text. Furthermore, the study by \citet{buschek21} indicates that offering suggestions provokes the use of more ``standard'' phrases (from a business email context).

Language models do not always have to generate text for direct integration in order to impact the writing. \citet{singh22} developed a writing tool that analyzes the current content of the text field and offers different kinds of matching stimuli, such as pictures, audio or text. Their user study shows that even without accepting the suggested text, this method helps the users when feeling stuck and inspires their subsequent writing.

Another related study shows that people find it easier to write themselves by observing generated text \cite{roemmele21}. In this process, before working on a task, the participants were shown how a system would solve this task with the help of text generation. This made it easier for the participants to solve the tasks independently, as they were shown a possible solution approach through the generation.

Two other examples, where generative models support writing without writing directly, are shown in the work of \citet{arnold21}. They proposed interaction methods which can be used to change the structure of a sentence in a meaningful way without distorting the basic statement. This approach can be used to get more variation into a longer text. Furthermore, the paper presents a technique that helps the user by formulating questions to be answered in the text in order to define guidelines for writing.

Overall, the related work shows that writing with AI can influence the text (e.g. wording and structure) but also impacts factors that go beyond (e.g. inspiration). These effects beyond text metrics motivate us to investigate another such factor in more detail, namely perceived authorship. Concretely, we compare perception and interaction when writing text with AI in two contrasting designs, that embed different human-AI roles, initiative, and control.

\section{Concepts and Design}

We designed the interactions and the UI of our prototypes in an iterative process and then decided on two final, contrasting interaction designs. 

\subsection{Design Process}

The aim of our design process was to identify interaction methods that have the potential to support mobile writing with the use of language models. In contrast to the methods presented above (see \Cref{sec:related_work}), where the AI is mostly in a helper role, we decided to explore designs with a more active appearance of the AI.
For example, the AI might write the text and the user would steer the AI and only edit the outcome when necessary. 

We started our design process with brainstorming and sketching in an expert group on possible interaction methods. We also took related work for inspiration into consideration \cite{arnold16, arnold20, buschek21, coenen21}.

Following an iterative approach, we conducted multiple rounds of sketching, discussions and collecting feedback among our research team over several weeks. 

Two key conceptual directions emerged: First, \wwss allows the user to choose the text to be integrated from a set of AI generated phrases. This represents the typical suggestion list design direction. Second, in \cgt,  the text is written ``live'' by the AI, and the user is supposed to only correct the text when it is going into a wrong direction. 

\begin{figure*}[t]
    \begin{center}
        \includegraphics[width=0.75\textwidth]{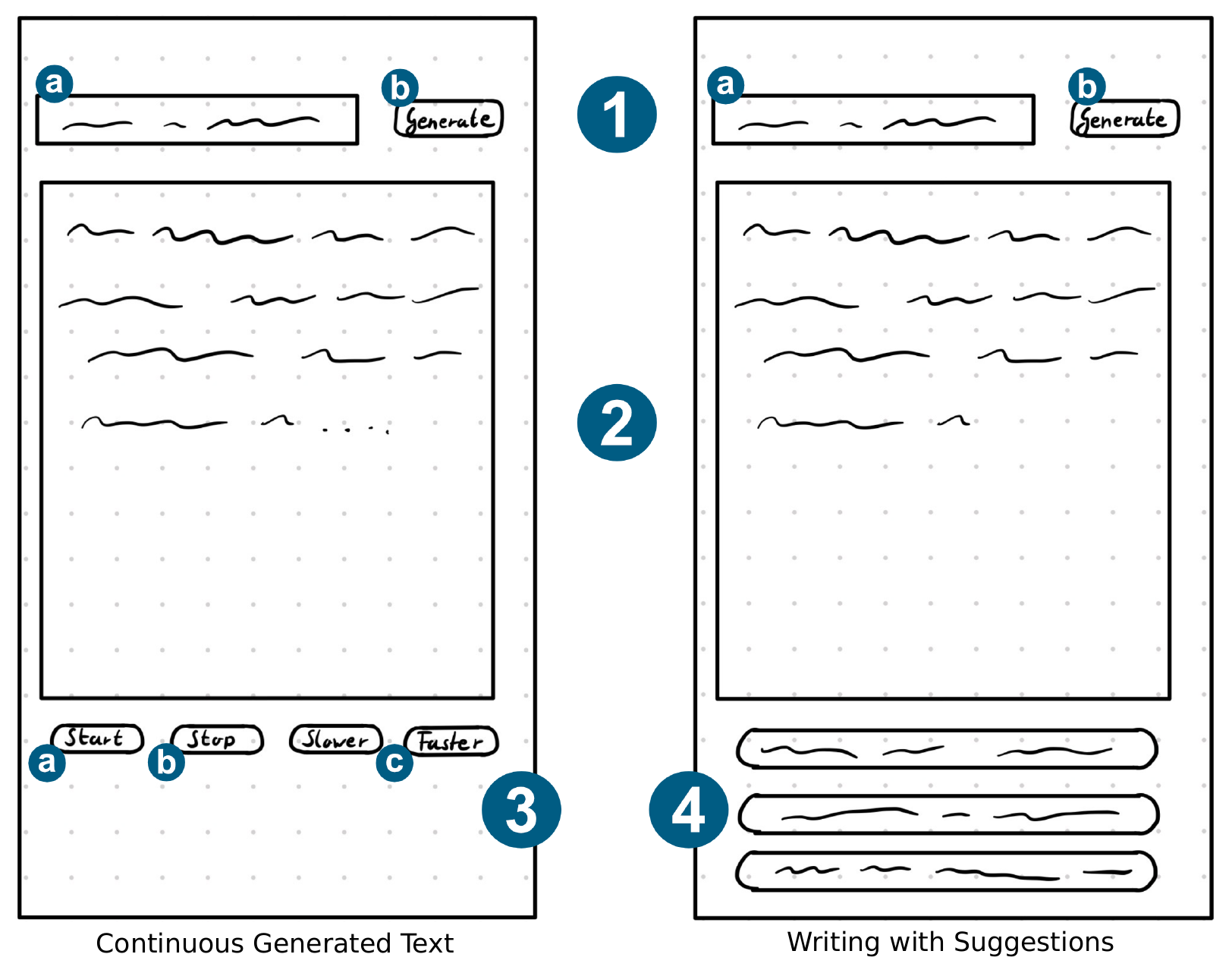}
        \caption{Early sketches of the developed concepts.}
        \Description{The left frame shows the interface for \cgt. The right frame shows the interface for \wws. In the top area, both UIs display a text input field next to a button to start generation. Below is a text field for generated text. The variants differ mainly in the bottom area. In the bottom area, \cgts provides buttons to control generation, \wwss displays three suggestions of generated phrases.}
        \label{fig:concepts}
    \end{center}
\end{figure*}

The main difference between these two methods is the human-AI relationship in terms of the distribution of ``roles'': In \cgt, the AI is the writer and the user acts as an editor. Instead, with \wws, the user actively controls the writing process by selecting from the supporting AI's suggestions. For both methods, we decided for an interaction where the user initiates the writing by entering text into a text field. This initial writing should indicate the topic of the text to be the starting text for the AI, but we ultimately removed this in favour of a typical single text entry area (more below). \Cref{fig:concepts} shows sketches of the two concepts.

Our concept and design phase was followed by an implementation and testing phase, as well as a small prestudy which resulted in the final versions of our prototypes. In this prestudy, the participants suggested improvements and were observed by the experimenter. 

\subsection{Final Designs}

Our final design decisions are explained in this section, along with a detailed description of the two implemented concepts. To explain our decisions, we refer to \Cref{fig:concepts} and recall the annotations of this figure to refer to distinct parts of the sketches.  

Based on our findings from the prestudy, we decided, for both methods, to remove the separate input field (1a) and combine its function with the main text field (2). Before this change, the user should provide the initial input in the input field (1a). With this change, however, the user simply enters the initial text directly in the main field (2).

\subsubsection{Continuous Generated Text}

A sketch of the early prototype UI \cgts is shown in \Cref{fig:concepts} on the left; the corresponding final version is shown in \Cref{fig:implementation} on the left.

The basic idea of this concept was that the user provides the input text and the AI continuously writes a text ``live'' wordy by word. The user can interrupt the AI while writing and edit the text, for instance, if it does not fit expectations.

The interaction starts with the user entering a sentence in the main text field (2). This starting text can be used to set the context and topic. After that, the AI starts writing: Its generated text appears over time, word by word, in the text field (2). At any time, the user can interrupt the AI by tapping the stop button (3b), for example, when the text deviates from the desired context or the user just wants to make manual improvements to the text. After stopping the AI writing, any parts of the text can be deleted or edited in the text field. When satisfied, the user is free to let the AI continue with writing by tapping on the start button (3a). This procedure can be repeated until the desired outcome is reached. 

The initial speed of the appearing text was implemented with one word per second. Thus, without intervention, the AI would write a text with 60 words per minute. That is close to twice as fast as the average typing speed on mobile devices \cite{palin19}. This should, in the best case, make writing with this interaction method more efficient while still allowing the user to potentially intervene in time.

Based on the feedback from the prestudy and our observations, we removed the buttons to adjust the AI writing speed (3c) and let the AI stop writing when the text field is selected by the user, in order to let the user edit the text.

\subsubsection{Writing with Suggestions}

\begin{figure*}[t]
    \begin{center}
    \includegraphics[width=0.75\textwidth]{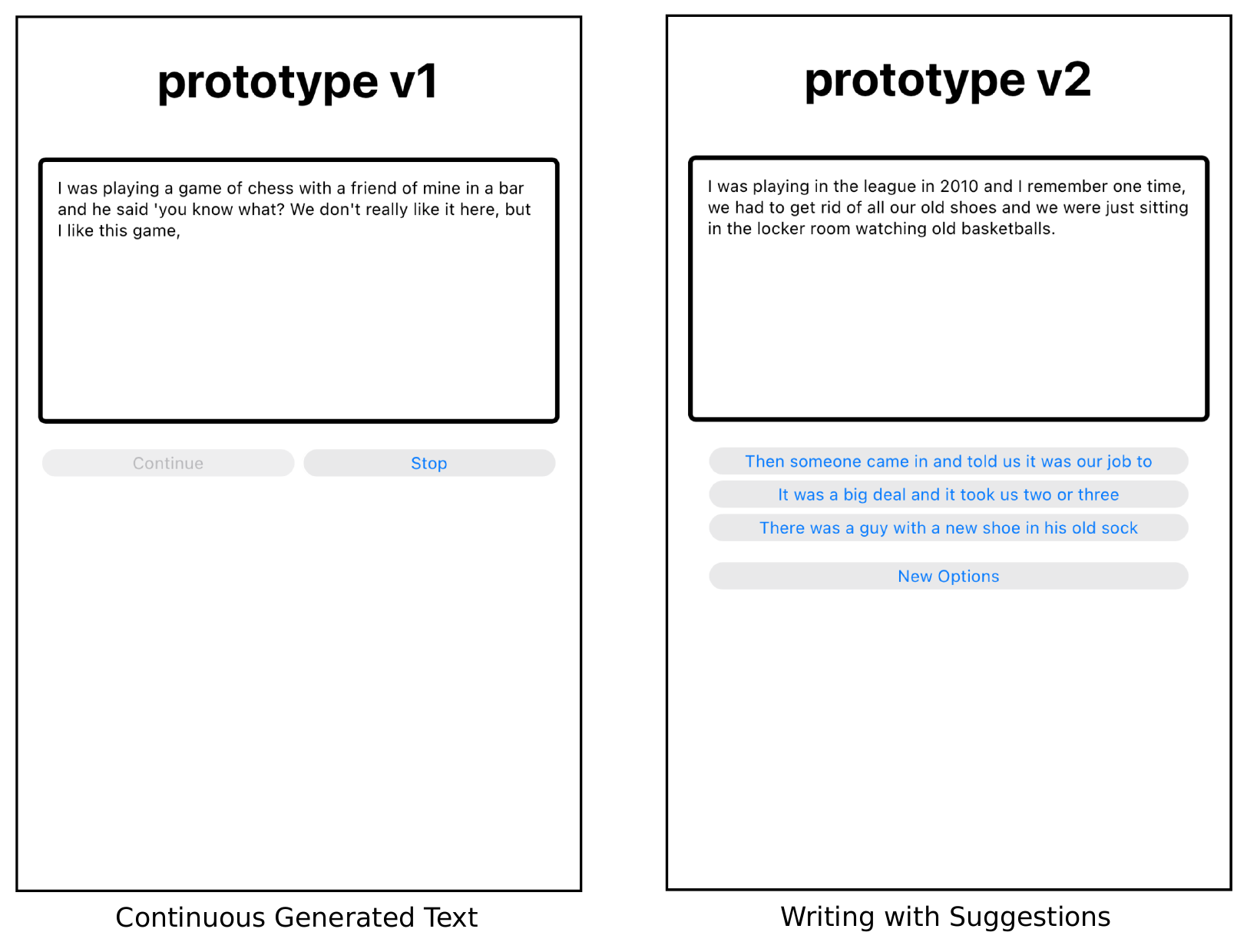}
    \caption{The final versions of the prototypes we implemented as mobile web-applications.}
    \Description{Both variants have a text input field in the top area. This text field includes text input and generated text. Below that, \cgts on the left shows a reduced set of buttons to start/continue and to stop the generation. \wwss on the right shows three text suggestions and a button to refresh suggestions.}
    \label{fig:implementation}
    \end{center}
\end{figure*}

Our early prototype for the method \wwss can be seen in \Cref{fig:concepts} on the right side. The corresponding final version can be seen in \Cref{fig:implementation}, also on the right.

The basic idea for this concept was that the user controls the text by selecting AI-generated phrases from a set of suggestions, similar to the use of suggestions in many such systems in related work. By using the language model, these suggestions are based on the text which has been already written.

Similar to \cgt, the user types an initial text into the main text field (2). The AI generates phrases and presents the suggestions below (4). When the user selects one of these suggestions, it is appended to the text and new suggestions are generated. We added a button below the suggestions to refresh the suggestions manually. This enables users to request new suggestions.

Based on the findings from the prestudy as well as our observations, we allowed the users to manually write and edit text in parallel to the suggestions, instead of writing solely with suggestions. To make the text interaction more natural, we added a delay of two seconds after the user stopped editing to let the AI present new suggestions.

The number of provided suggestions was informed by related work \cite{buschek21}. This paper shows that with displaying more suggestions, inspiration is promoted, but at the expense of efficiency. We decided to display three suggestions at the same time to balance between efficiency and inspiration. We limited the AI-generated phrases to a length between eight and twelve words. If a phrase completes a sentence, any words behind a terminal punctuation (point, question mark, exclamation mark) are cut off. 

\section{Study}

\subsection{Design}
We used a $3 \times 2$ within-subject design to compare three interaction methods in two tasks each. The methods were: \cgt, \wws, and a baseline (manual typing, no AI). We used two writing tasks (see \Cref{subsec:task}). The combination and order of the writing tasks and the methods was counterbalanced following a Latin-Square \cite{gao05}.

We logged interactions during writing as follows: For each text we recorded the submitted result, the needed time, the number of generations used and for \wwss also how often a user refreshed the given suggestions. In addition, any interaction with the text, such as typing or deleting a character, was logged along with a timestamp. We also video-recorded most of the sessions with participants' consent.

\subsection{Participants}
A total of 18 people (14 male, 4 female) participated in the study. All participants were recruited via social networks or email lists.
Their age ranged from 18 to 33 years, with a median of 24 years. Originally, we had 20 participants in our study, yet two dropped off due to technical reasons.

All had an English level of A1\footnote{CERF scale: \url{https://www.coe.int/en/web/common-european-framework-reference-languages/table-1-cefr-3.3-common-reference-levels-global-scale} \lastaccessed} or higher and were therefore familiar with the English language.
One person was left-handed, the rest were right-handed. Regarding writing preferences, 15 participants reported holding the smartphone with both hands and to use both thumbs for writing, the remaining three preferred to type one-handed with the thumb.

On the question if they had previous experiences with generated text, 14 people answered with yes. Most referred to using word suggestions or auto-correction, on their mobile devices (eleven people). A few participants had already gained experiences with generative writing, for example, when composing mails with Gmail or intelligent translators (three people). 

Participants were compensated with €10.

\subsection{Apparatus}
Based on our concepts, two prototypes were realized as mobile web-applications for the study. The frontend was implemented with JavaScript, using the UI library React\footnote{\url{https://www.reactjs.org} \lastaccessed}. For the backend, Python was used together with the web framework Flask\footnote{\url{https://flask.palletsprojects.com/en/2.0.x} \lastaccessed}. Text generation was realized with the pre-trained language model GPT-2. \Cref{fig:implementation} shows the final UIs.

The prototypes were deployed online. This way, the participants could easily complete all tasks on their own smartphone. Participants took part in the study via a provided link, without having to install software beforehand.

\subsection{Task}\label{subsec:task}
Participants had the task to write different short texts on mobile devices. We designed two tasks to have a wider range of situations to test the interaction methods: The Birthday-Task is more specific and constraint through exact instructions on what to write. In contrast, the Vacation-Task is more open.
We set the minimum length of a text to 20 words, but had no constraints on the maximum length. The texts had to be written in English.

\begin{table}[h]
\begin{tabularx}{\columnwidth}{@{}l X}
    Birthday-Task: & \textit{It is your friend's birthday. Write an email to wish all the best and mention that you have to} \\ & \textit{meet again in some time.} \\
     
    Vacation-Task: & \textit{Write a short story about your last or upcoming vacation.}
\end{tabularx}
\end{table}

\subsection{Procedure}
The study was conducted as a supervised online study via video calls. It was separated into three parts. The procedure started with an introduction to the study in line with our institutional procedures. It was followed by introducing the interaction methods and the task briefing. To complete the procedure, we presented a post-hoc questionnaire in the final step. Throughout the process, an experimenter was present to respond to questions or problems. Verbal communication happened mostly in German, with exceptions where the participants preferred English. The complete procedure took 35 to a maximum of 80 minutes, with an average duration of about 50 minutes.

\subsubsection{Introduction}
At the beginning, the participants were informed about the study and the data protection guidelines, which had to be agreed in order to take part. The two prototypes, \cgts and \wws, were then presented and explained to the participants. In order to get a feeling for the interaction with the prototypes, the participants were allowed to test the two methods voluntarily before starting with the tasks.

\subsubsection{Main part}
We briefed the participants to complete both tasks, as presented in \Cref{subsec:task}. For each task, three texts had to be written. One text in each of the two prototypes and one manually with a standard text input for reference. In total, the participants had to write six texts. After completing a text, the participants rated six statements, for example on satisfaction and authorship, on a five-point Likert scale. An overview of the questions is shown in the table in the appendix (\Cref{subsec:appendix-likert}).

\subsubsection{Questionnaire}
After completing all six possible combinations of tasks and methods, participants filled in a post-hoc questionnaire. It asked questions on socio-demographics, the interaction methods, and general feedback. 
An overview of the questions of the post-hoc questionnaire can be seen in the tables in \Cref{subsec:appendix-questionnaire}.

\section{Results}

To analyze our data, we adapted similar approaches from related work \cite{buschek21, wobbrock16}. We processed the data with Python\footnote{Python, \url{https://www.python.org/} \lastaccessed} and analyzed it with R\footnote{R: A Language and Environment for Statistical Computing, \url{https://www.r-project.org/} \lastaccessed}. We used R mainly for significance testing through linear mixed-effect models (LMMs) and relied for that on the package lmerTest \cite{lmerTest}. For the LMMs, the factors \textit{subject} (participant) and \textit{task} were taken into account as random effects, the only fixed effect was the \textit{input method} (interaction method). Based on the LMMs, post-hoc pairwise comparisons were computed with a Tukey test (Z-test) and a Bonferroni–Holm correction applied. For the ordinal Likert data, we applied the Friedman test and the Wilcoxon signed-rank test with the package coin \cite{coinPackage}. First, we ran a Friedman test to check for general differences in data. If differences were found, we subsequently ran Wilcoxon signed-rank tests to compare the data pairwise. The Wilcoxon signed-rank test was computed with Bonferroni correction. The significance threshold in our reports is $p < .05$. 

Throughout the results, we use the abbreviations \stdiabbrs for baseline input (manual standard text input), \cgtabbrs for \cgts, and \wwsabbrs for \wws.

\subsection{Dataset Overview}

The produced texts contained overall, 6699 words with a mean of 62.03 (SD 29.55). Each text was written within a mean of 3.04 minutes (SD 1.82 minutes), excluding the time needed for generating text. The system generated text within a mean of 3.14 seconds (SD 2.08). A total of 987 generations were executed in the conditions \cgts and \wws, with a mean of 13,71 (SD 11.90). New suggestions were requested with a mean of 5.14 times (SD  5.96) in the \wwss condition.

\begin{figure*}
    \centering
    \begin{subfigure}[b]{0.45\linewidth}
        \subcaption{Words per minute}
        \label{fig:result_textmetrics_wpm}
        \includegraphics[width=\linewidth,trim=0 5mm 0 13mm, clip]{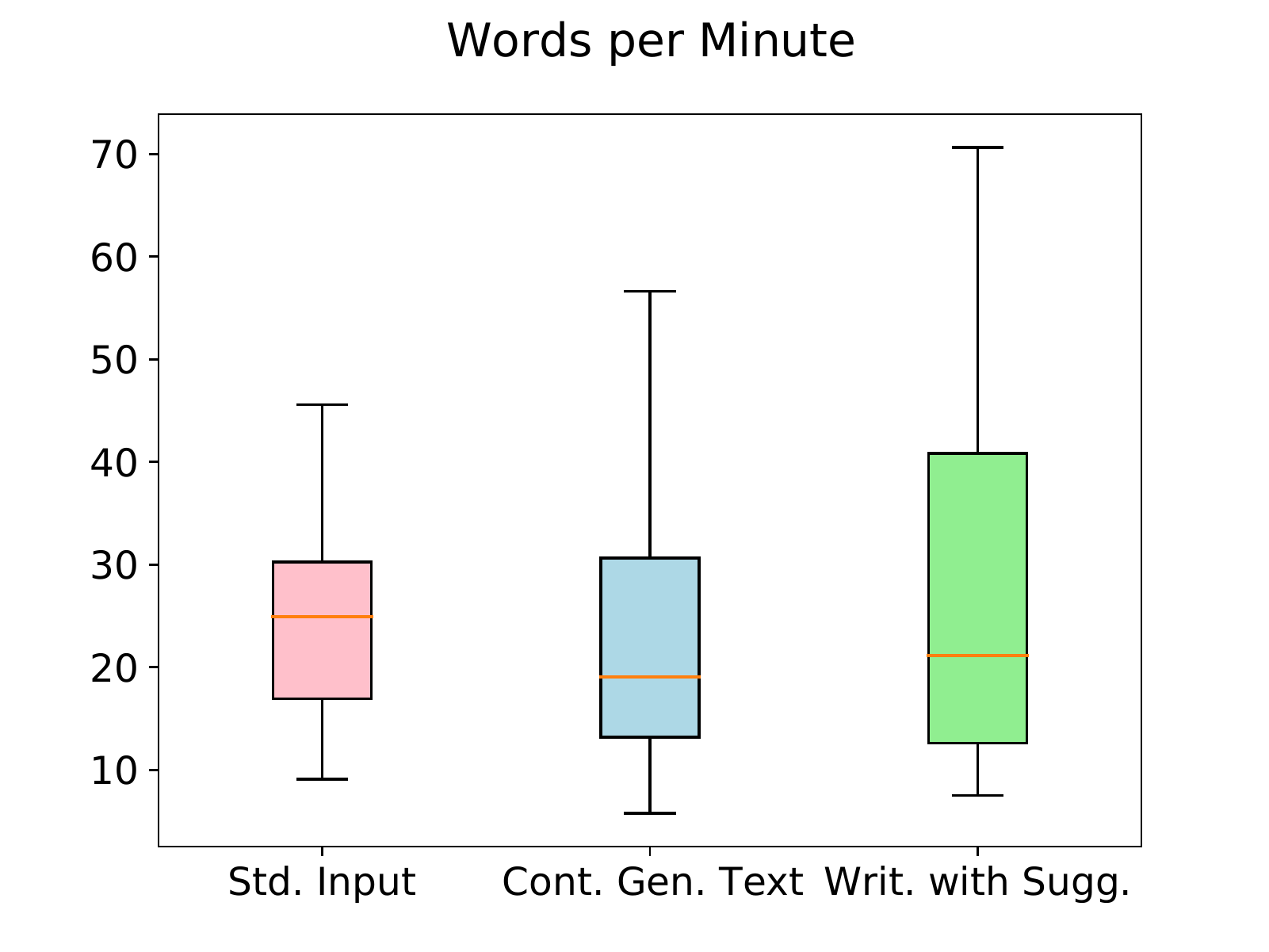}
    \end{subfigure}
    \hspace{10pt}
    \vspace{10pt}
    \begin{subfigure}[b]{0.45\linewidth}
        \subcaption{Text length (characters)}
        \label{fig:result_textmetrics_textlength}
        \includegraphics[width=\linewidth,trim=0 5mm 0 13mm, clip]{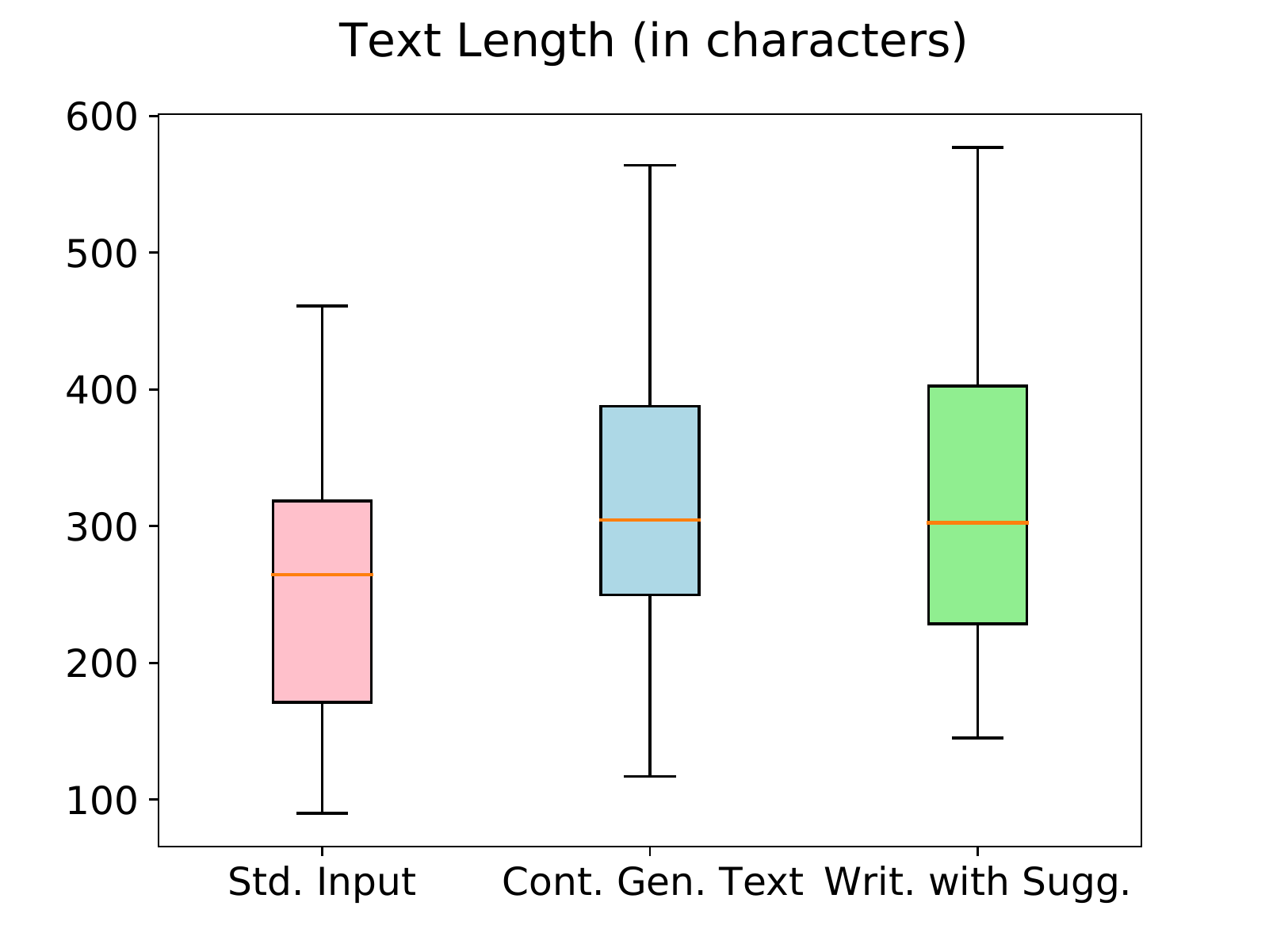}
    \end{subfigure}
    \begin{subfigure}[b]{0.45\linewidth}
        \subcaption{Task completion time (seconds)}
        \label{fig:result_textmetrics_taskcompletiontime}
        \includegraphics[width=\linewidth,trim=0 5mm 0 13mm, clip]{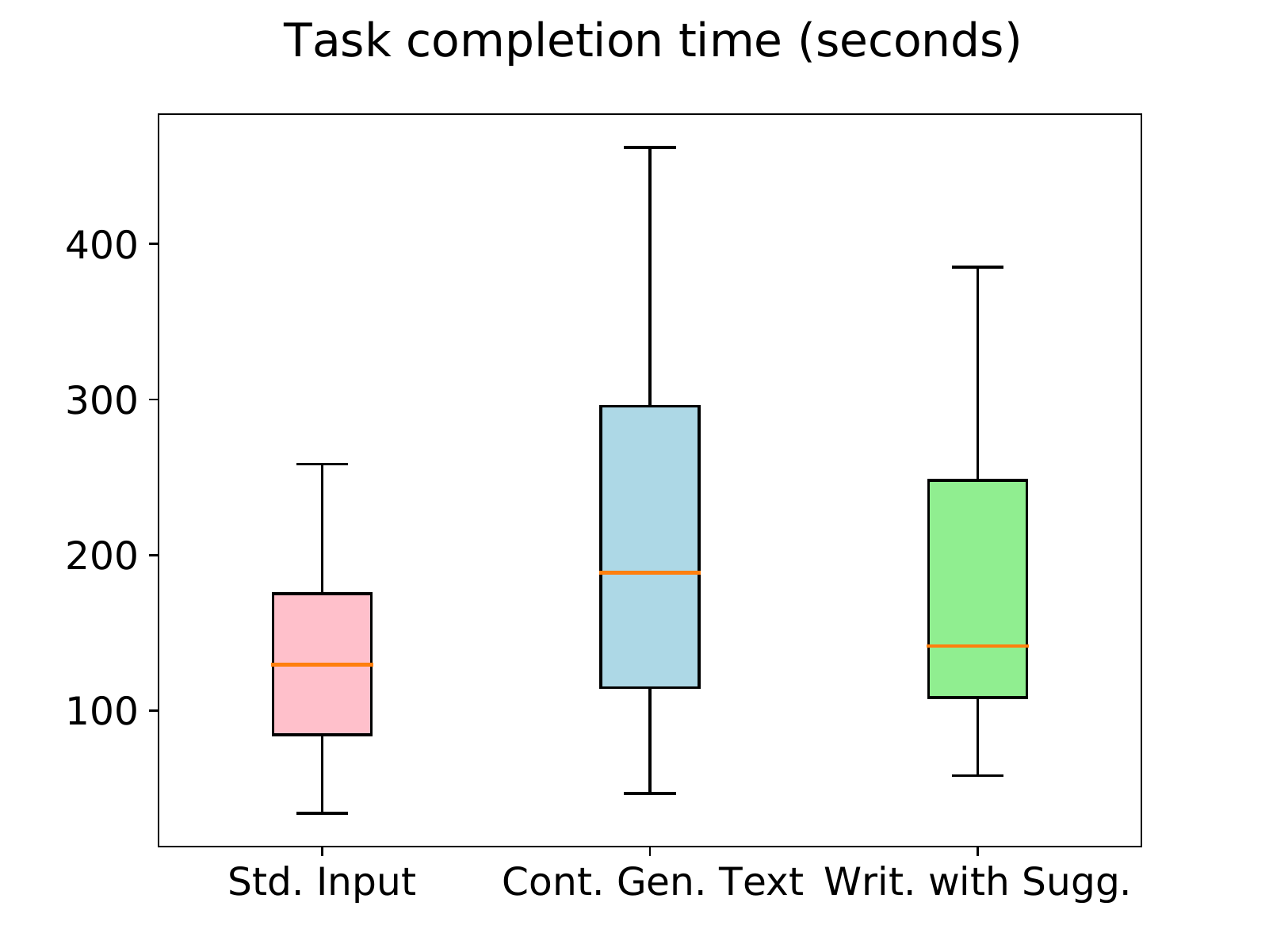}
    \end{subfigure}
    \caption{Visualized results of input performance and text metrics a) Words per minute (WPM), b) text length measured in characters, c) task completion time measured in seconds (excluding the time the system needed for generation).}
    \Description{WPM across conditions differs only slightly but \wwss shows the highest deviation. Text length is longer with the generative interaction methods. Task completion time is longer with the generative interaction methods, and \cgts shows the highest deviation.}
    \label{fig:result_textmetrics}
\end{figure*}

\subsection{Words per Minute and Text Length}\label{subsec:wpm_textlength}

We computed the word count for the words per minute measure by dividing the character length by five, as suggested by \citet{arif09}. This result was then divided by the minutes, people needed to absolve the task.


As can be seen in \Cref{fig:result_textmetrics_wpm} WPM across conditions differed only slightly (\meandesc{\stdiabbr}: 25.36 WPM, \meandesc{\cgtabbr}: 23.02 WPM, \meandesc{\wwsabbr}: 27.61 WPM), differences were non-significant. 






Text length was measured in characters. The LMMs had both generative interaction methods as positive predictors, \cgtabbr: \glmmci{78.17}{27.095}{24.42}{131.91}{= .005} and \wwsabbr: \glmmci{80.08}{27.095}{26.34}{133.83}{= .004}. 

Pairwise comparisons were only statistically significant between the generative interaction methods and \stdiabbr, \ztest{\cgtabbr}{\stdiabbr}{2.89}{= .01}, \ztest{\wwsabbr}{\stdiabbr}{2.96}{= .009}. Writing with generative models produced longer text (\meandesc{\stdiabbr}: 259.64 chars, \meandesc{\cgtabbr}: 337.81 chars, \meandesc{\wwsabbr}: 339.72 chars). 

\subsection{Task completion}

We measured the task completion time in seconds from displaying a prototype until completing the task. The time, the system needed to generate the text, was subtracted from the task completion time.






The descriptive overview shows that writing with generative models takes more time than without (\meandesc{\stdiabbr}: 146.26 s, \meandesc{\cgtabbr}: 211.94 s, \meandesc{\wwsabbr}: 189.04 s).
Results from LMMs showed the input method to be a significant and positive predictor for task completion time in \cgtabbr: \glmmci{65.68}{18.18}{29.62}{101.74}{< .001} and \wwsabbr: \glmmci{42.78}{18.18}{6.72}{78.85}{= .02}.

Pairwise comparisons were only statistically significant between the generative interaction methods and \stdiabbr, \ztest{\cgtabbr}{\stdiabbr}{3.61}{< .001}, \ztest{\wwsabbr}{\stdiabbr}{2.35}{= .049}. 

\subsection{Text Input and Corrections}

We measured the keystroke events to analyze the text input and corrections. Since both measures were counts, we fitted generalized LMMs (Poisson family) on them, short GLMMs. We further computed chances of a method influencing text input or corrections for positive $\beta$ values ($exp(\beta)$) and negative $\beta$ values ($1 - exp(\beta)$).

\subsubsection{Text Input}\label{subsec:results_text_input}

To analyze text input, we rely on the counts of keyboard strokes per person.

Average keystrokes showed to be minimally higher in \cgtabbrs and lower in \wwsabbrs compared to \stdiabbrs (\meandesc{\stdiabbr}: 325.08 chars, \meandesc{\cgtabbr}: 341.25 chars, \meandesc{\wwsabbr}: 170.91 chars).

The GLMMs had method as a significant predictor. It was a positive predictor for \cgtabbr: \glmmci{.05}{.01}{.02}{.07}{< .001} and a negative predictor for \wwsabbr: \glmmci{-.64}{.02}{-.67}{-.61}{< .001}. The chance of executing more keystrokes is increased by 5.13\% for \cgtabbr. While the chance is decreased by 47.2\% for \wwsabbr. 

All pairwise comparisons were statistically significant, \ztest{\stdiabbr}{\cgtabbr}{3.758}{< .001}, \ztest{\wwsabbr}{\stdiabbr}{-40.833}{< .001}, \ztest{\wwsabbr}{\cgtabbr}{-40.833}{< .001}.







\begin{figure*}
    \centering
   \begin{subfigure}[b]{0.45\linewidth}
        \subcaption{Number of input actions (keystrokes)}
        \label{fig:result_inputs_inputactions}
        \includegraphics[width=\linewidth,trim=0 5mm 0 13mm, clip]{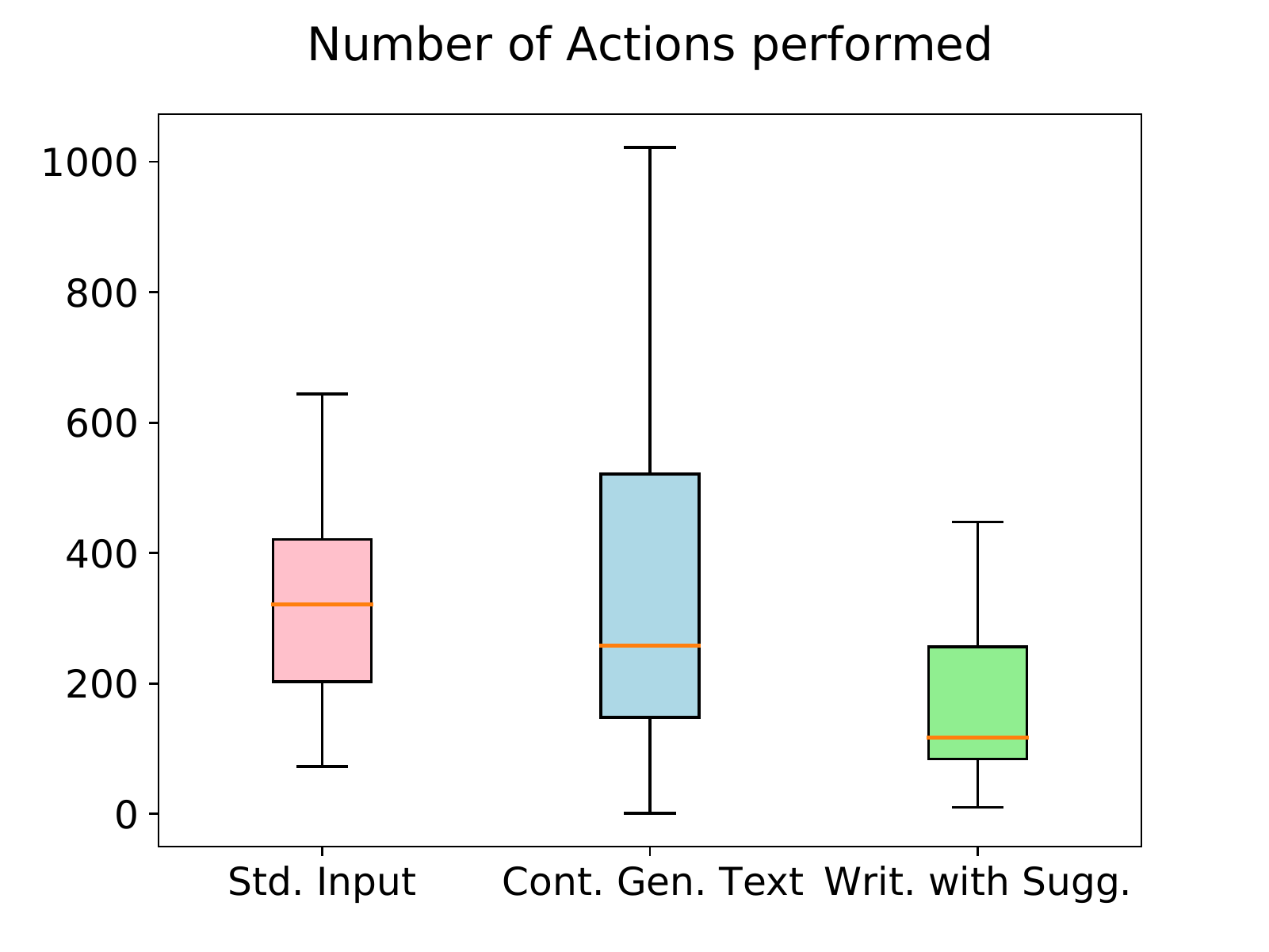}
    \end{subfigure}
    \hspace{10pt}
    \vspace{10pt}
    \begin{subfigure}[b]{0.45\linewidth}
        \subcaption{Number of backspaces}
        \label{fig:result_inputs_numbackspaces}
        \includegraphics[width=\linewidth,trim=0 5mm 0 13mm, clip]{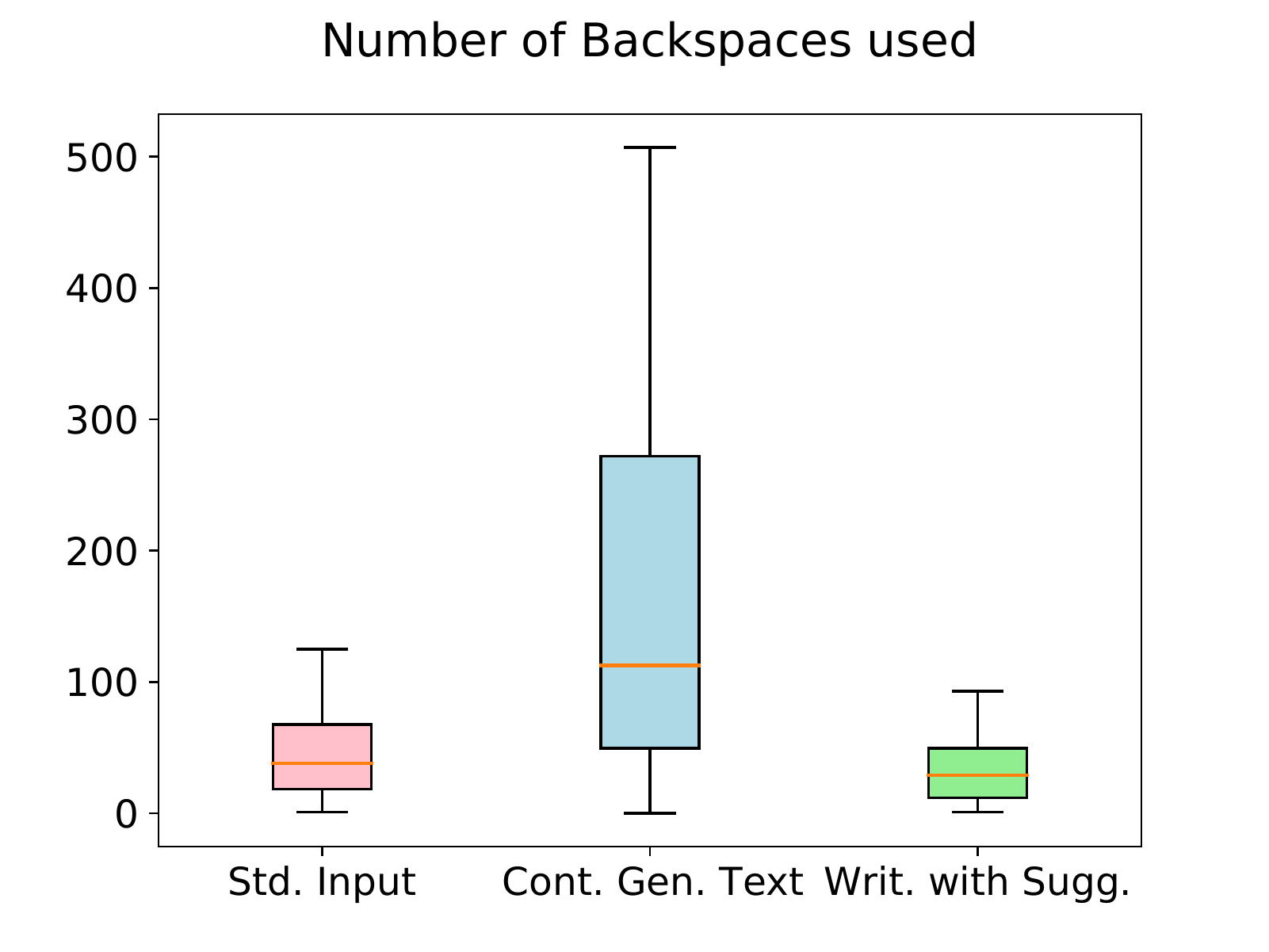}
    \end{subfigure}
    \begin{subfigure}[b]{0.45\linewidth}
        \subcaption{Number of backspace sequences}
        \label{fig:result_inputs_numbackspacesequences}
        \includegraphics[width=\linewidth,trim=0 5mm 0 13mm, clip]{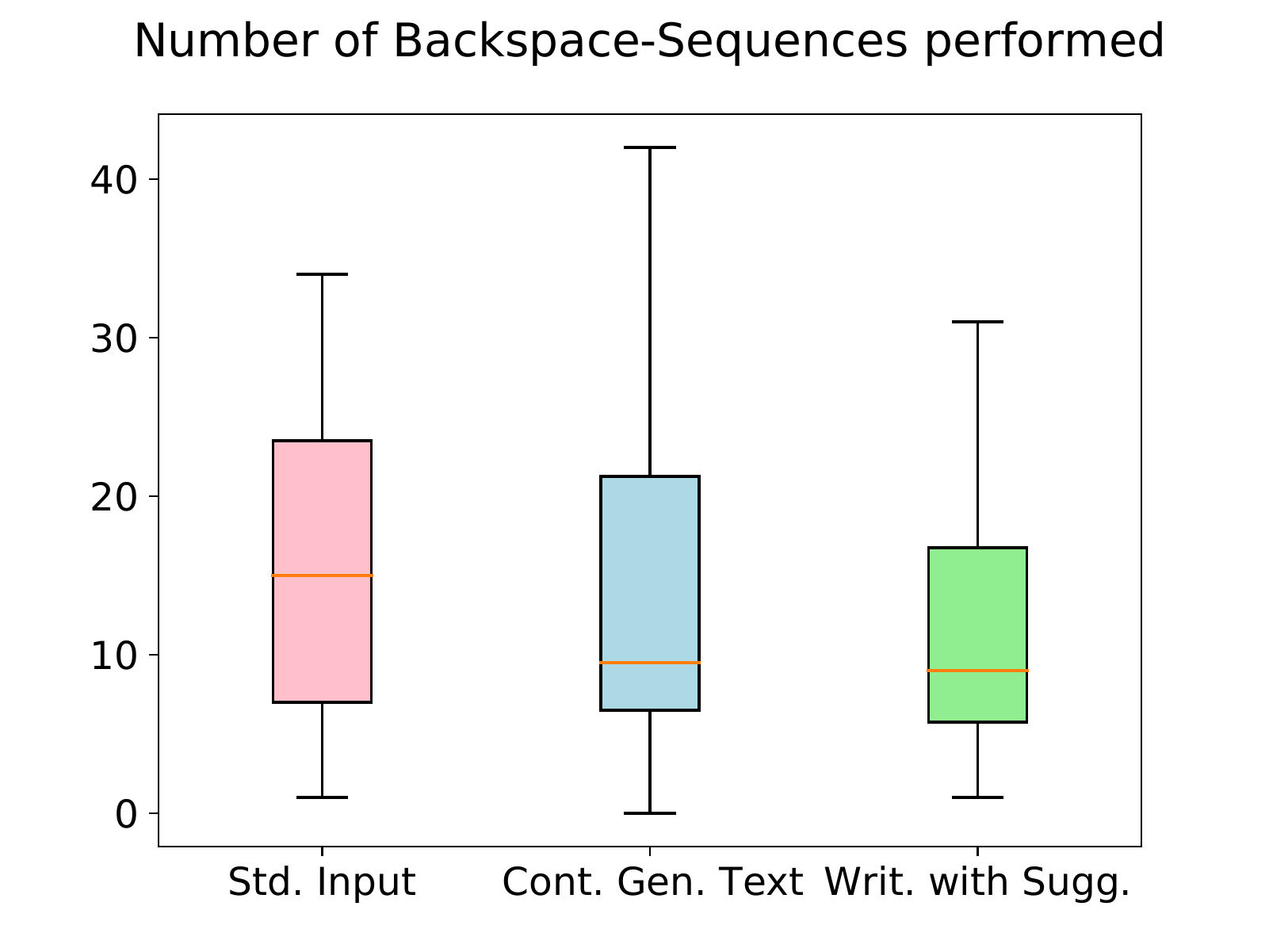}
    \end{subfigure}
    \begin{subfigure}[b]{0.45\linewidth}
        \subcaption{Length of backspace sequences}
        \label{fig:result_inputs_lengthbackspacesequences}
        \includegraphics[width=\linewidth,trim=0 5mm 0 13mm, clip]{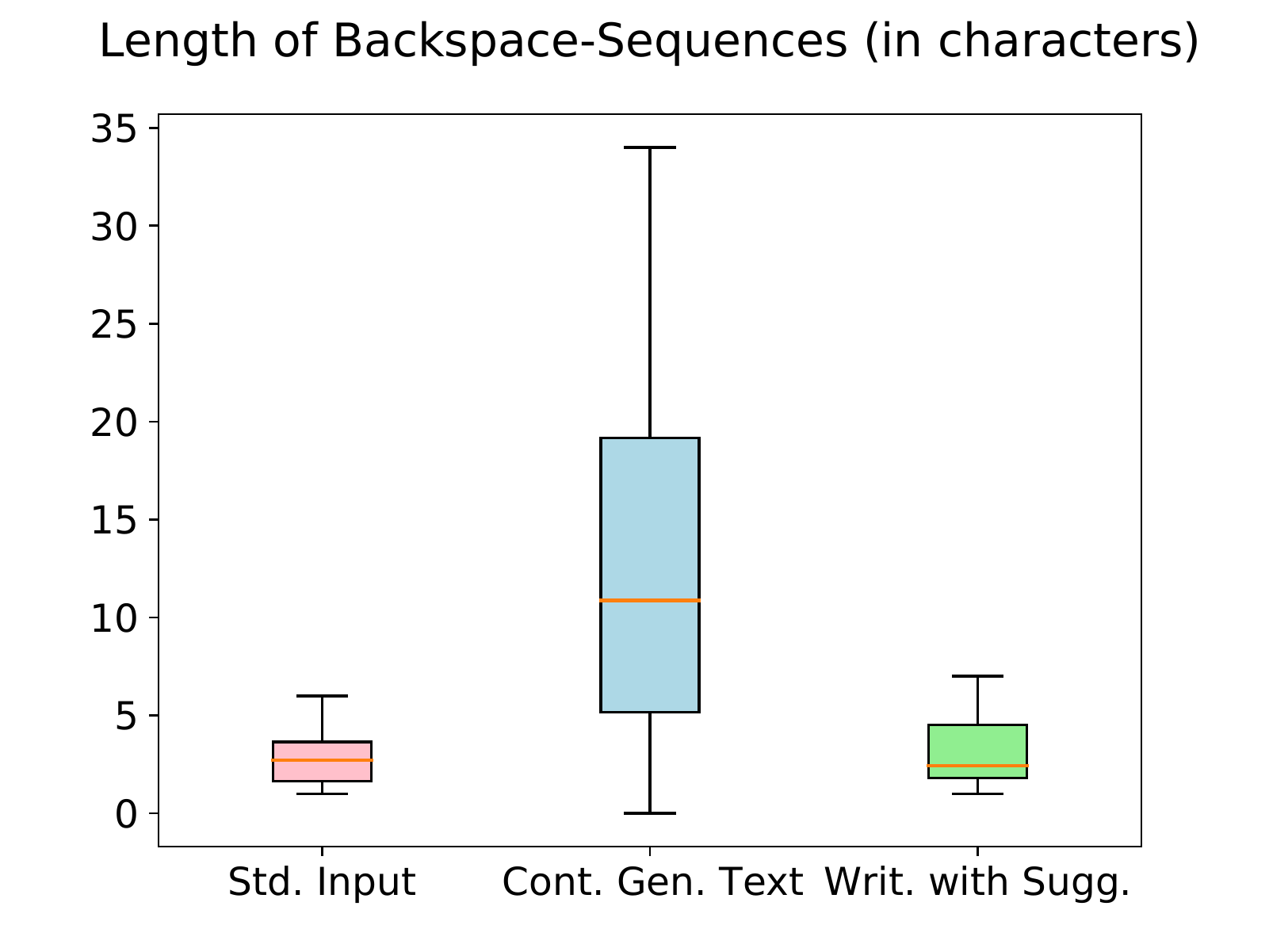}
    \end{subfigure}
    \caption{Visualized results of the users' text interactions. a) Number of input actions based on all keystrokes, b) number of backspaces for the purpose of correction, c) number of backspace sequences where a sequence is longer than one backspace, and d) the average length of backspace sequences.}
    \Description{The number of input actions is the lowest for \wws. \cgts shows the highest deviation. The number of backspaces is the highest for \cgts and lowest for \wws. \cgts shows the highest deviation. The number of backspace sequences is lower for generative interaction methods. The average length of backspace sequences is the longest for \cgts with a high deviation.}
    \label{fig:result_inputs}
\end{figure*}

\subsubsection{Corrections}\label{subsec:results_text_correction}





%

To analyze corrections, we took exclusively the backspace keystrokes into consideration. The descriptive statistics show heavily increased corrections (backspaces) for \cgtabbrs and slightly decreased corrections for \wwsabbr. (\meandesc{\stdiabbr}: 45.89, \meandesc{\cgtabbr}: 187.83, \meandesc{\wwsabbr}: 39.67). The GLMMs, we fitted on the data, had method as a significant predictor. It was a positive predictor for \cgtabbr: \glmmci{1.41}{.03}{1.36}{1.46}{< .001} and a negative predictor for \wwsabbr: \glmmci{-.15}{.04}{-.22}{-.07}{< .001}. \wwss decreased the chances for text correction by 14\%, whereas the chances are almost four times higher (410\%) when writing with \cgt. 

The pairwise comparisons were statistically significant for all cases, \ztest{\stdiabbr}{\cgtabbr}{51.376}{< .001}, \ztest{\wwsabbr}{\stdiabbr}{-4.035}{< .001}, \ztest{\wwsabbr}{\cgtabbr}{-53.421}{< .001}.

In addition, we analyzed backspace sequences. A backspace sequences is defined as multiple backspace keystrokes that happen subsequently. Descriptively, the number of backspace sequences was reduced for generative methods compared to \stdiabbrs and the lowest overall for \wwsabbrs (\meandesc{\stdiabbr}: 15.86, \meandesc{\cgtabbr}: 13.83, \meandesc{\wwsabbr}: 12.39). For the number of backspace sequences, the GLMMs had method as a significant predictor. It was a negative predictor for \cgtabbr: \glmmci{-.14}{.06}{-.26}{-.02}{= .025} as well as for \wwsabbrs: \glmmci{-.25}{.06}{-.37}{-.12}{< .001}. \cgts decreased the chances for backspace corrections by 13\%, with \wwss chances are decreased by 22\%.

The pairwise comparison was nearly statistically significant for \ztest{\cgtabbr}{\stdiabbr}{-2.24}{= .065} and statistically significant for \ztest{\wwsabbr}{\stdiabbr}{-3.92}{< .001}.






Moreover, we ran an analysis on the averaged backspace sequence length. Descriptive statistics show a heavily increased sequence length (\meandesc{\stdiabbr}: 3.04, \meandesc{\cgtabbr}: 13.05, \meandesc{\wwsabbr}: 3.60). The LMMs we fitted on the data had method as a significant and positive predictor for \cgtabbr: \glmmci{10.01}{1.50}{7.04}{12.98}{< .001}.

In a pairwise comparison, only the comparison to \cgtabbrs were statistically significant, \ztest{\cgtabbr}{\stdiabbr}{6.68}{< .001}, \ztest{\wwsabbr}{\cgtabbr}{-6.31}{< .001}. More text needs to be corrected with \cgt.







\subsection{Perception}
\begin{figure*}
    \centering
    \begin{subfigure}[b]{0.47\linewidth}
        \subcaption{The interaction method was suitable for this task.}
        \label{fig:likert_ratings_suitabiliy}
        \includegraphics[width=\linewidth,trim=0 15mm 65mm 24mm, clip]{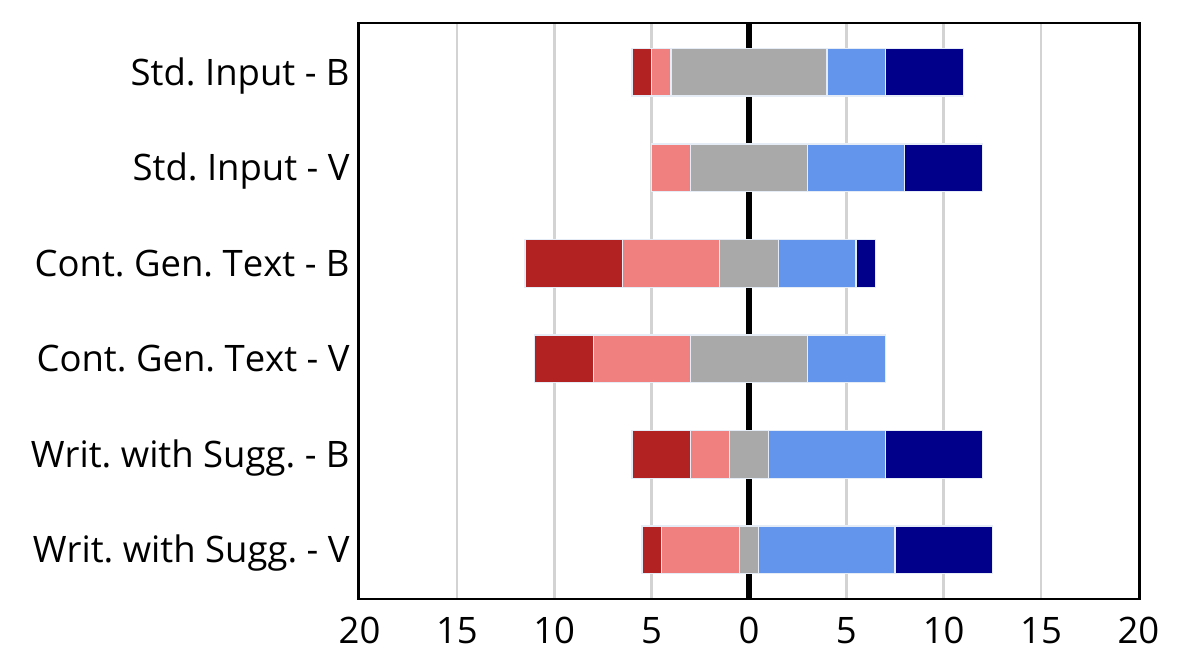}
    \end{subfigure}
     \hspace{5pt}
    \vspace{8pt}
    \begin{subfigure}[b]{0.47\linewidth}
        \subcaption{The interaction method helped me writing the text.}
        \label{fig:likert_ratings_helpfulnes}
        \includegraphics[width=\linewidth,trim=0 15mm 65mm 24mm, clip]{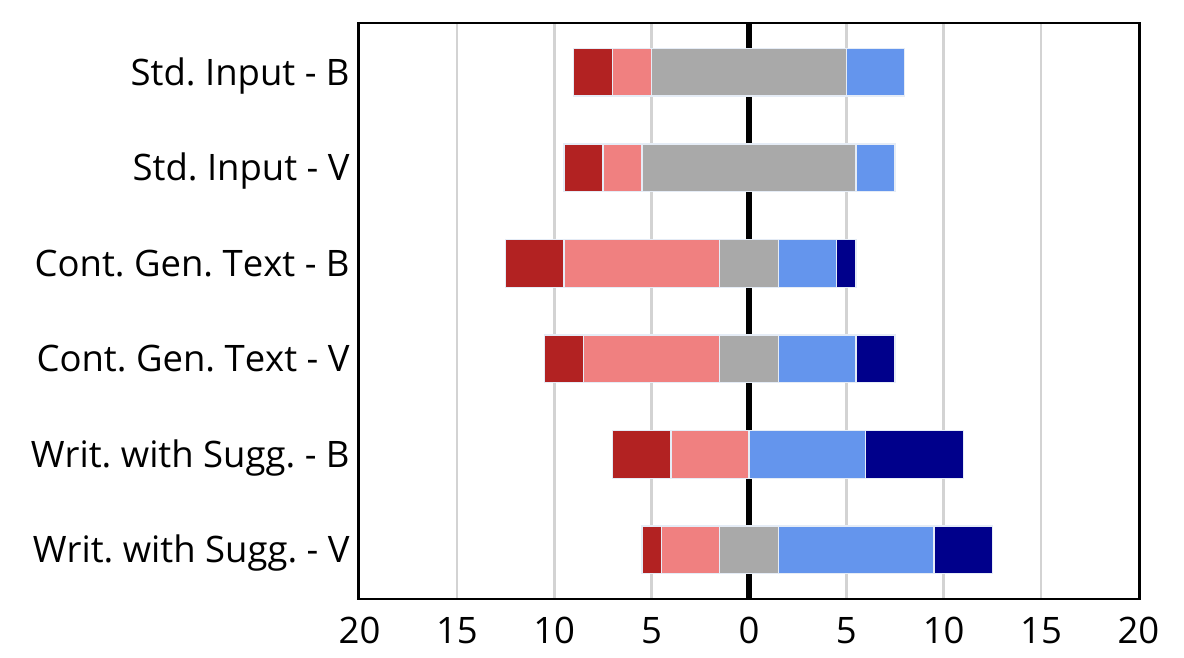}
    \end{subfigure}
   \begin{subfigure}[b]{0.47\linewidth}
        \subcaption{I am satisfied with the text.}
         \label{fig:likert_ratings_satisfaction}
         \includegraphics[width=\linewidth,trim=0 15mm 65mm 24mm, clip]{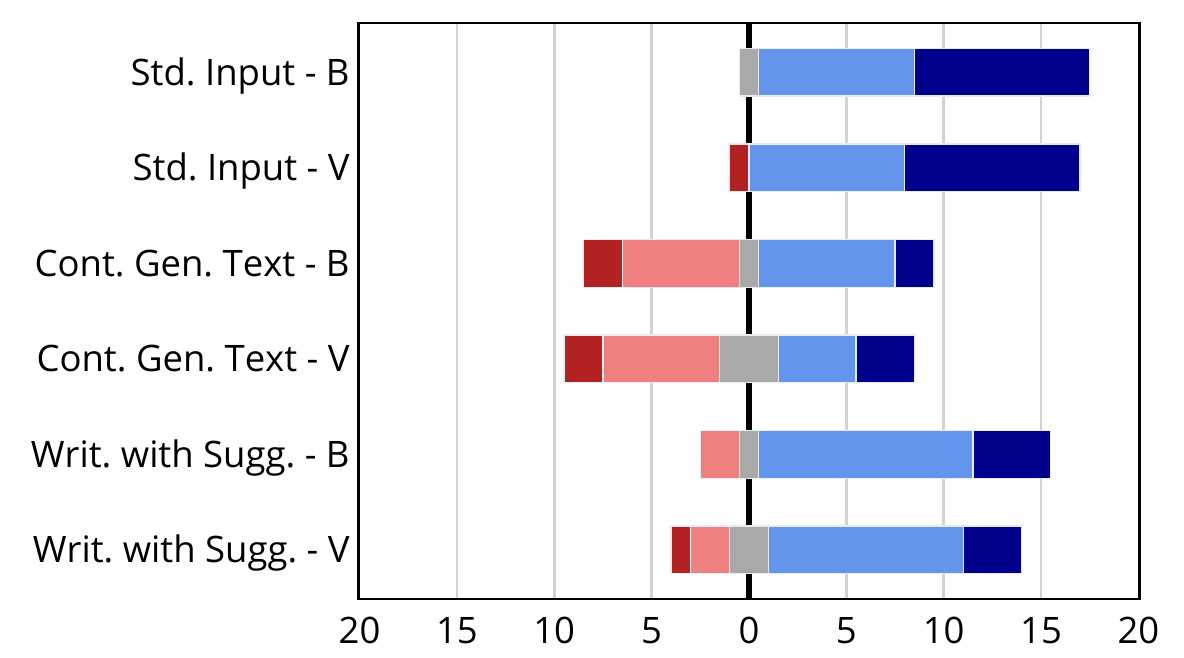}
    \end{subfigure}
    \hspace{5pt}
    \vspace{8pt}
    \begin{subfigure}[b]{0.47\linewidth}
        \subcaption{It was easy for me to write the text.}
        \label{fig:likert_ratings_difficulty}
        \includegraphics[width=\linewidth,trim=0 15mm 65mm 24mm, clip]{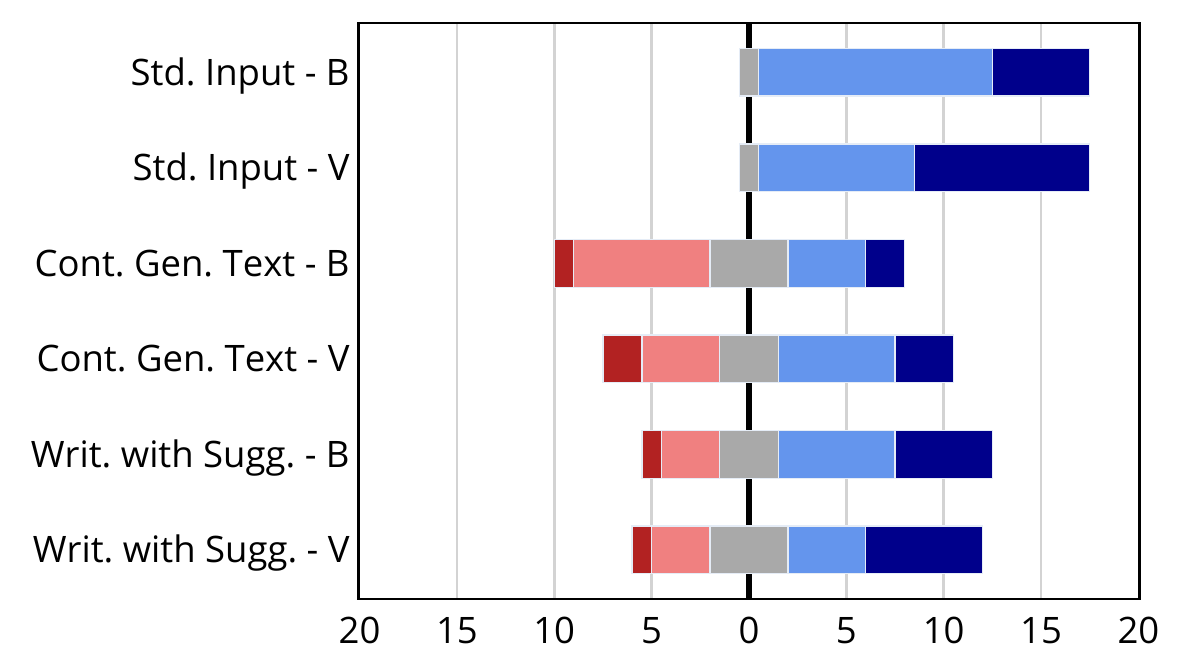}
    \end{subfigure}
   \begin{subfigure}[b]{0.47\linewidth}
        \subcaption{I feel like I am the author of the text.}
        \label{fig:likert_ratings_authorship}
        \includegraphics[width=\linewidth,trim=0 15mm 65mm 24mm, clip]{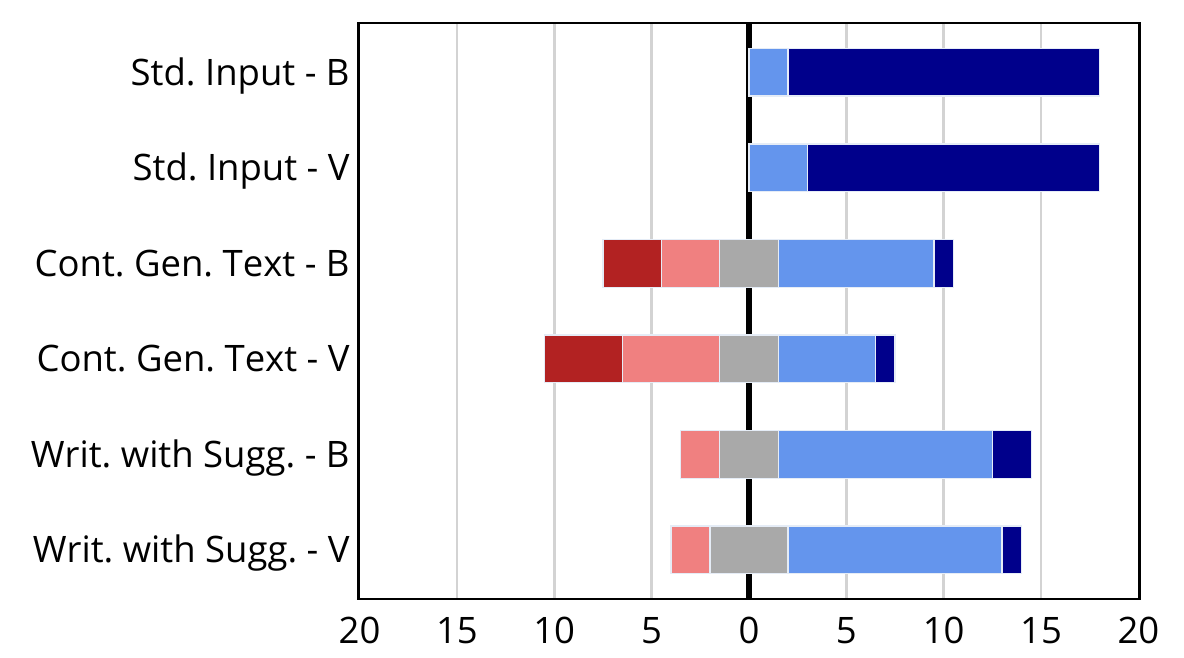}
    \end{subfigure}
    \hspace{5pt}
    \vspace{8pt}
    \begin{subfigure}[b]{0.47\linewidth}
        \subcaption{The interaction method influenced the wording of the text.}
        \label{fig:likert_ratings_wording}
        \includegraphics[width=\linewidth,trim=0 15mm 65mm 24mm, clip]{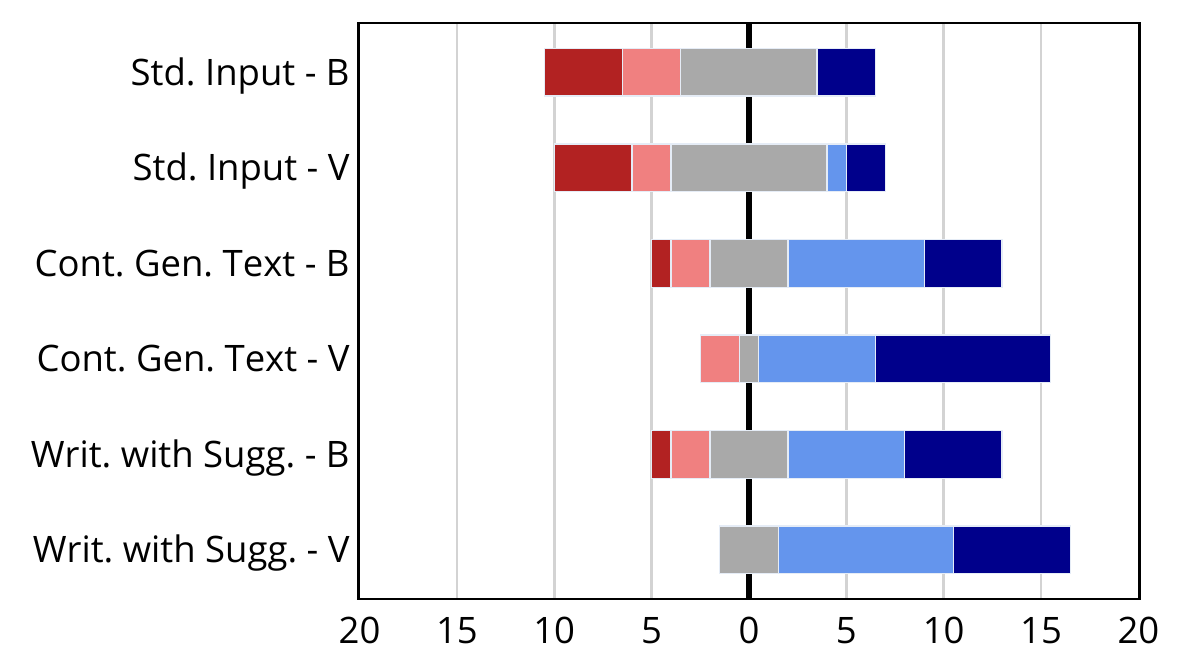}
    \end{subfigure}
    \begin{subfigure}[b]{0.65\linewidth}
        \includegraphics[width=\linewidth,trim=0 0 0 0, clip]{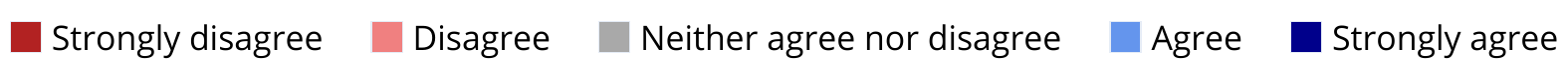}
    \end{subfigure}
    \caption{Overview of Likert ratings from the questionnaires we presented to the user after each task. For each method we display two bars: The abbreviation B stands for Birthday-Task, V for Vacation-Task. The x-axes display absolute participant numbers. The legend is displayed below the charts.}
    \Description{Chart a) shows ratings on suitability. People rated \wwss more suitable than \cgt. Chart b) shows ratings on helpfulness. \wwss was rated positively, and \cgts negatively. Chart c) shows ratings on satisfaction. The largest part rated positively in \wwss, and ratings in \cgts were divided almost equally. Chart d) shows ratings on ease of use. In comparison to the base condition, both generative interaction methods were rated as less easy. Chart e) shows ratings on perceived authorship. The largest part of people writing with \wwss perceived themselves as authors. In \cgt, fewer people perceived themselves as authors. Chart f) shows ratings on the methods' influence on the wording. The generative interaction methods were both rated to influence wording by the largest part of people.}
    \label{fig:likert_ratings}
\end{figure*}

People rated their perception on each task via Likert items (see \Cref{fig:likert_ratings}). Ratings were given on a scale from one to five, from ``strongly disagree'' to ``strongly agree''. For the sake of our analysis, we concentrated on comparing the methods and therefore average the ratings per person per method. These averaged ratings result in a scale of ten steps instead of five steps. Since only two values were averaged, the mean is equal to the median. The order of the ratings is maintained. To analyze the data, we first ran a Friedman test to check for differences between ratings on methods. When the test result was significant, we ran a Wilcoxon signed-rank test with Bonferroni-Holm correction.

\subsubsection{Suitability and Helpfulness}

The medians for ratings on suitability were \stdiabbr: 3.5 (IQR = 1), \cgtabbr: 2.5 (IQR = 1.75), \wwsabbr: 3.5 (IQR = 1.375). These differences were statistically significant \fmann{18}{9.5}{< .01}. Pairwise comparison was only significant for \cgtabbrs vs. \wwsabbrs (\wilcox{-2.82}{= .009}). \wwss was perceived to be more suitable to write text on mobile devices in comparison to \cgt.

Ratings on helpfulness showed medians as follows, \stdiabbr: 3 (IQR = 0.5), \cgtabbr: 2.5 (IQR = 1.375), \wwsabbr: 3.5 (IQR = 1.375). The test on differences was statistically significant \fmann{18}{7.40}{< .05}. Again, only the comparison between \cgtabbrs and \wwsabbrs were 
significant in pairwise comparisons (\wilcox{-2.74}{= .013}). Compared to \cgt, \wwss was perceived as more helpful when writing text on mobile devices. 




%
%







\subsubsection{Satisfaction and Difficulty}\label{subsec:results_satisfaction_difficulty}

The ratings on satisfaction show medians as follows, \stdiabbr: 4.5 (IQR = 1), \cgtabbr: 3.0 (IQR = 1.75), \wwsabbr: 4.0 (IQR = 1). The test on differences was statistically significant \fmann{18}{16.91}{< .001}. All pairwise comparisons were significant, \stdiabbrs vs. \cgtabbrs: \wilcox{3.10}{= .003}, \stdiabbrs vs. \wwsabbr: \wilcox{2.63}{= .027}, \cgtabbrs vs. \wwsabbr: \wilcox{-2.41}{< .043}. Text written without generative models satisfied people the most. Text written with \cgts was perceived as least satisfying.

Median ratings on difficulty were \stdiabbr: 4.5 (IQR = .5), \cgtabbr: 3 (IQR = 1.25), \wwsabbr: 3.5 (IQR = 1.875). The differences were statistically different \fmann{18}{13.13}{< .002}. In pairwise comparisons, only the comparison \stdiabbrs vs. \cgtabbrs was significant (\wilcox{3.28}{< .001}).













\subsubsection{Authorship and Wording}\label{subsec:results_authorship_wording}

For the ratings on authorship the medians were as follows, \stdiabbr: 5 (IQR = .375), \cgtabbr: 3 (IQR = 1.375), \wwsabbr: 4 (IQR = 0.875). The test for these differences was significant \fmann{18}{33.63}{< .001}. All pairwise comparison showed statistical significance, \stdiabbrs vs. \cgtabbr: \wilcox{3.73}{< .001}, \stdiabbrs vs. \wwsabbr: \wilcox{3.76}{< .001}, \cgtabbrs vs. \wwsabbr: \wilcox{-3.23}{< .001}. When using \wwss most people still perceived themselves as author. They perceived less authorship when writing with \cgt.

Ratings on how the method influenced the wording of the text showed descriptively following differences, \stdiabbr: 3 (IQR = 1.75), \cgtabbr: 4 (IQR = 1), \wwsabbr: 4 (IQR = 1). These differences were statistically significant \fmann{18}{11.29}{=.004}. Within the pairwise comparisons, only the comparisons \stdiabbrs vs. \cgtabbrs (\wilcox{-2.95}{= .006}) and \stdiabbrs vs. \wwsabbrs (\wilcox{-2.89}{= .01}) were statistically significant. Both generative interaction methods were perceived by people to influence the wording of the text.













\subsection{Subjective Feedback}
\begin{figure*}
    \centering
    \begin{subfigure}[t]{0.48\linewidth}
    \vspace{10pt}
        \subcaption{How hard was it to learn and use this method?}
        \label{fig:questionnaire_usability}
        \includegraphics[width=\linewidth,trim=22mm 18mm 28mm 27mm, clip]{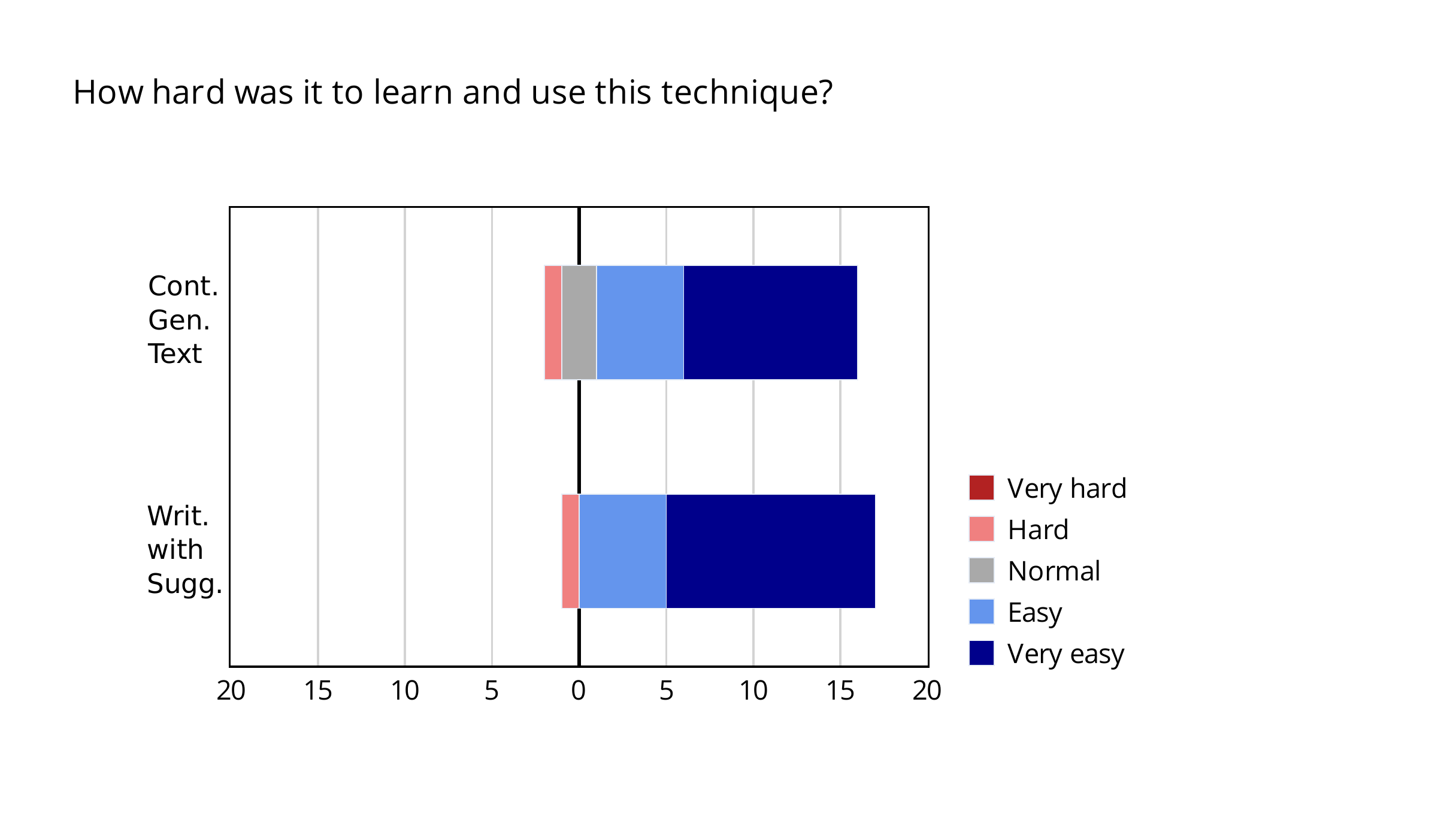}
    \end{subfigure}
    \begin{subfigure}[t]{0.48\linewidth}
        \subcaption{The interaction method inspired me to write things I normally would not have thought of.}
        \label{fig:questionnaire_inspiration}
        \includegraphics[width=\linewidth,trim=22mm 18mm 28mm 27mm, clip]{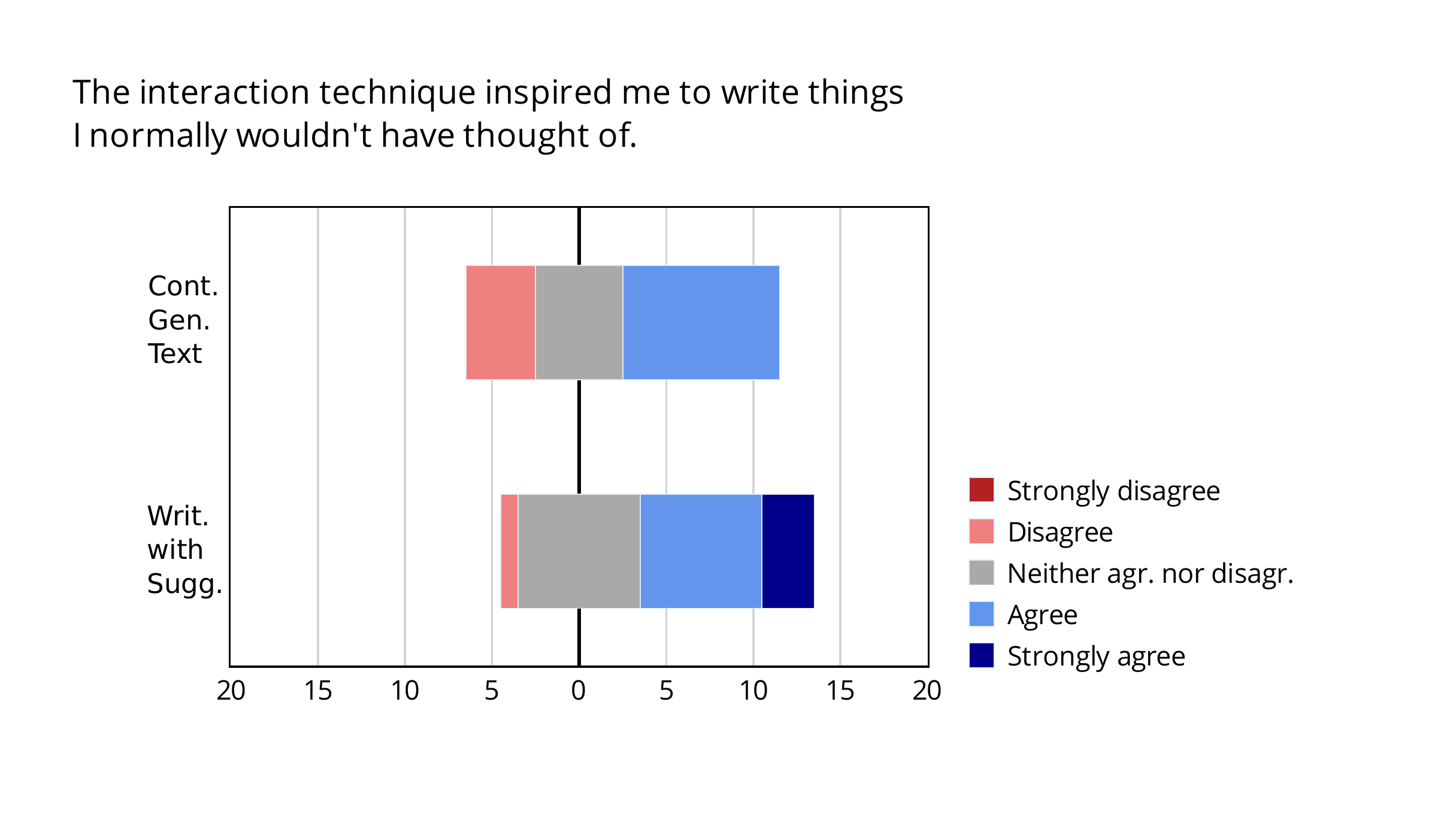}
    \end{subfigure}

    \caption{Results of the post-hoc questionnaire on a) usability as a combination of how hard it was to learn and use a method and b) inspiration. X-axes reflect the absolute numbers of participants.}
    \Description{Almost all people rated the generative interaction methods to be easy or very easy to learn. People mainly agree on being inspired by both methods.}
    \label{fig:questionnaire}
\end{figure*}

In a post-hoc questionnaire, people rated further aspects on Likert scales and provided feedback on open-ended questions. 
We asked them to rate usability, specific aspects of the generative interaction methods, e.g. generation speed in the condition \cgt, and inspiration in the condition \wws. Regarding inspiration, we asked them two questions, but dismissed one questions since the question contained speculations about the future. All ratings we consider for this analysis are visualized in \Cref{fig:questionnaire}.

\subsubsection{Usability and Inspiration}\label{subsec:results_usability_inspiration}

Most of people found both generative interaction methods easy or very easy to learn and use. Easy was rated by five people in each condition (27.78\% \cgtabbrs and \wwsabbr). Very easy was rated by ten people (55.56\%) for \cgtabbrs and by twelve people (66.67\%) for \wwsabbr.

To assess how much people felt inspired by the generative interaction methods, we asked them to provide ratings on the item ``the interaction method inspired me to write things I normally would not have thought of''. For \cgt, nine people (50\%) agreed on that. For \wws, seven people agreed (38.89\%) and three people strongly agreed (16.67\%). Both methods can inspire people, however, giving suggestions is more suitable for inspiration.

\subsubsection{Ratings of the Methods}

\begin{figure*}
    \centering
    \includegraphics[width=0.75\linewidth]{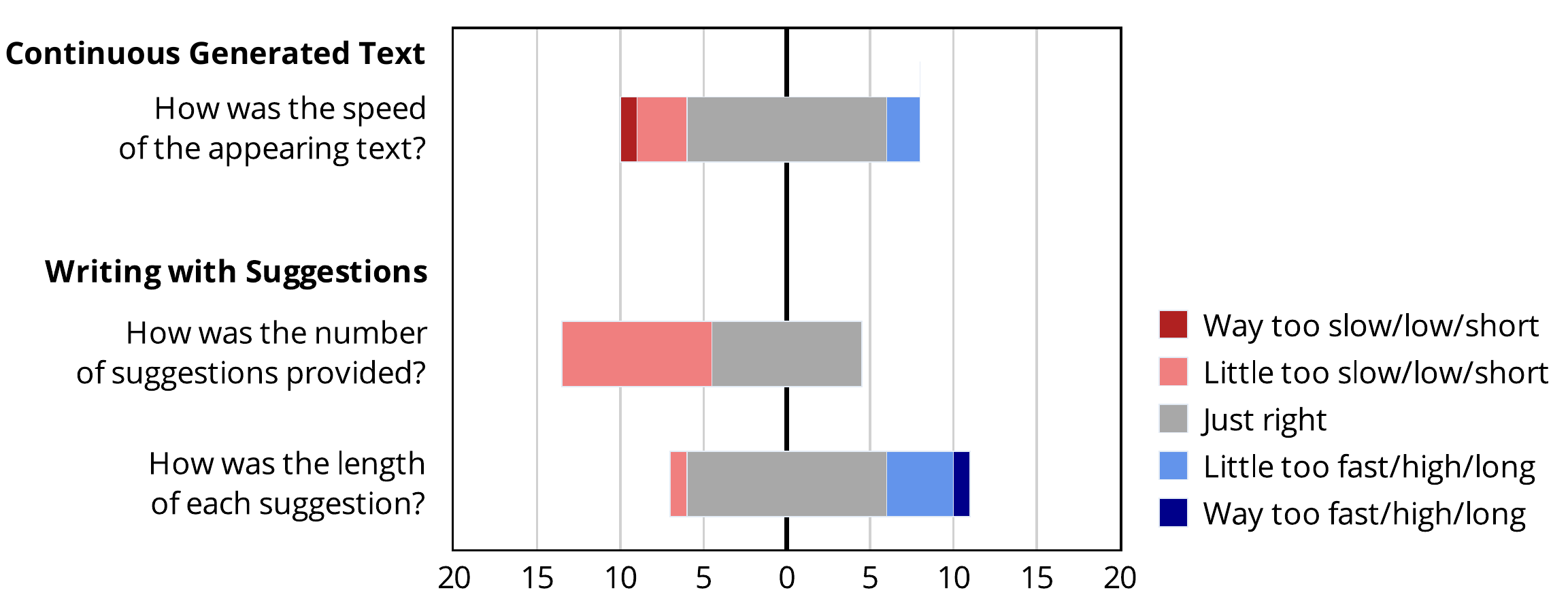}
    \caption{Visualization of method specific results from the post-hoc questionnaire. The top bar shows ratings on the AI writing speed within method \cgt ( legend: ``too slow'' to ``too fast''). The lower bars show ratings on the number of suggestions (legend: ``too low'' to ``too high'') and length of the suggestions (legend: ``too short'' to ``too long'') within the method \wws. The x-axes reflect absolute participant numbers.}
    \Description{For \cgt, most people found the speed of the appearing text just right. For \wws, almost half of people rated the number of suggestions as too low but found the length of each suggestion to be just right, with a light tendency of too long suggestions.}
    \label{fig:combined_ratings}
\end{figure*}

We also asked people to rate specific attributes regarding the generative interaction methods. Visualizations can be seen in \Cref{fig:combined_ratings}.

For the method \cgtabbrs we asked how people perceived the generation speed. The majority, twelve people (66.67\%), found the generation speed just right. Only a small fraction, three people (16.67\%) found it a little too slow, and only one person (5.56\%) rated for way too slow. In general, the generation speed as sufficient, but could be accelerated a little. 

For the method \wwsabbrs we asked how the number and length of suggestions was perceived. Regarding the number of suggestions, the replies were two-fold. Nine people (50\%) found them too less, while the other nine people (50\%) found the number of suggestions to be just right. Thus, the number of suggestion should be increased to turn out just right for the majority of users.

The length of suggestions was rated by most people as just right. This rating was given by twelve people (66.67\%). Four people (22.22\%) found the length a little too long, and only one person found the length way too long.

\subsubsection{Feedback on Birthday Task}
To write a birthday greeting, 14 people (77.78\%) preferred the method \wwss over \cgt. People argued to have a better control over the text (13), to be more efficient (3), and found suggestions inspiring (2). 

However, there were only three people that would prefer \wwss also over a standard text input. 
The 14 people (77.78 \%) who preferred the standard input over any method working with generated text, found it easier to integrate personal information (8) and mentioned a better control over the text (7). 
One person who preferred one of the generational methods reasoned: ``I find it hard to come up with new ideas for everyone when greeting them on their birthdays. The generated text provides a source of inspiration, even when the wording has to be changed by hand.''. Besides being inspired from the methods (2) another argument for preferring generational methods was an improved efficiency when writing texts (2).

\subsubsection{Feedback on Vacation Task}
To write a story about a vacation, twelve people (66.67\%) preferred the method \wwss over \cgt. One participant, which preferred \wws, stated: ``The Continuous Generated Text produced more relevant sentences than [in the other task], but still was suggesting things that I didn't really want to tell, in the story. So I prefer Writing with Suggestions, because at least I could somehow control where the story is going.''
The reasons for preferring \wwss were therefore mainly the better control of the text (7) and the more precise text that could be produced as a result (7).

A total of seven persons (38.89\%) stated to prefer one of the generative methods over the standard text field, from which three preferred \cgts and four \wwss. The reasons for preferring these methods were mainly the inspiration (3) and efficiency (2).
The eleven persons preferring the standard text field argued that, when using the standard text field, the text is more precise and personal (6) and it is easier and faster to write the text (3). 

\subsubsection{Feedback on the Method Continuous Generated Text}
It was suggested that  \cgts could be useful for longer texts (4). In particular, for standard texts, like emails, (7) or some sort of formal texts (7). These are text types that have often a common structure and so give a good opportunity to be automated.
Another suggestion that was made are fictional texts (4), where the method has a lot of freedom and also can give a lot of inspiration for writing.

As advantages of \cgts were mentioned efficiency (7), in order to automate special sorts of text, and inspiration (11), for content and wording of the text.

The main disadvantage of \cgts is the lack of control over the text (12). This makes it hard for users to include special facts into the text or express exactly what someone wants to say (8). These factors make the method also a bit inefficient to use (4), which is expressed by the answer of one user: ``It's slow in the sense that you have to wait quite long before you can see the result. Also long passages have to be deleted by hand, which is also slow and potentially error prone.''

According to the users, one feature, which would improve the method, is an optimized language model for creating better text (7). Moreover, they suggested allowing to input special keywords or themes in order to gain more control over the text (7). One participant described these keywords as words that should definitely be included in the text so that they can serve as some sort of guideposts.

\subsubsection{Feedback on the Method Writing with Suggestions}
The participants mentioned that \wwss can be useful in a lot of different situations (4), especially when writing longer texts (5). Similar to \cgt, standard texts (9) and emails or chat-messages (9) were frequently mentioned as good use cases.

The big advantage of \wwss were the suggestions. They gave the users the option to always choose the phrases that fitted the most (6), which mostly resulted in useful text (5). Moreover, through the use of suggestions, the users found this method efficient and comfortable (10). Another advantage was frequently mentioned considering suggestions to be inspiring for new topics and different wording (8).

As disadvantages, participants mentioned that the suggestions sometimes were unusable (4), which resulted in long searching times for the right suggestion (5). 

For this method, the participants also suggested an optimized language model as a feature. In the best case, this should be user-specific (3) in order to generate better and more suitable suggestions (4). Furthermore, it was mentioned that the suggestions should be shorter (2) and more different (1).

\section{Discussion}

We discuss differences between the two interaction methods we proposed for writing text on mobile devices. These differences relate mainly to the aspects of control and initiative. Moreover, we discuss efficiency, inspiration and roles. Furthermore, we highlight the perceived influence of the interaction methods on wording and authorship. Finally, we give design recommendations and note research implications.

Overall, we only looked at writing short texts on mobile devices. Findings might deviate for other use cases, for example, other interaction methods or other language models. 

\subsection{Comparison of the Methods: Control and Initiative Differs, Generated Suggestions Decrease Text Input}

With our experiment, we have shown that the design of human-AI interaction influences the level of control and initiative. We have tested two interaction methods in this regard. These methods clearly differ in the way how the generated text is displayed and integrated.

Differences in control and initiative within the presented interaction methods are reflected in the interaction data. Writing with \cgts increased manual input, while it was heavily decreased for \wwss (see \Cref{fig:result_inputs_inputactions} \&  \Cref{subsec:results_text_input}). With the method \cgt, the system writes text live word by word. Writing text with \cgts shifts the focus of the user from writing to editing, and user and system switched initiative in turn. In comparison, with the method \wws, we presented three generated completions of sentences to the user. The user was free to decide to take over one of these suggestions. In parallel, text could be manually written by the user. Only the user had the initiative of writing text. The user had full control over the adoption of a particular generated text.

These differences in initiative and control are also visible in the data on text correction. Text written with \cgts needs correction more often, and backspace sequences are longer than in \wwss or writing manually (see \Cref{fig:result_inputs_numbackspaces}, \Cref{fig:result_inputs_lengthbackspacesequences} \& \Cref{subsec:results_text_correction}). If text gets automatically written by the system, but does not correspond to the user's notion, these generations have to be removed completely up to the point where the user finds the text appropriate. The user has to take back the initiative and exerts control, mostly by editing and correcting AI text. This is prevented in \wwss since the user can filter the generated text upfront and decides to take over the suggestion. Thus, with \wws, the user has the initiative and keeps control through selecting suggestions.

According to the literature on mixed-initiative interaction \cite{horvitz99}, we contribute two  practical examples for mixed-initiative interaction with generative AI. At the same time we challenge the term mixed-initiative since Horvitz defined initiative only implicitly through interfaces that enable efficient collaboration between agent and user. We suggest considering initiative more concretely as a ``design material'', that is, a variable to explicitly manipulate and reason about when designing such mixed-initiative interfaces. More broadly, we argue that design of intelligent systems should at least consider giving the system the initiative to manipulate content directly to be a truly mixed-initiative design approach. Following this definition, only \cgts was a mixed-initiative interface, since the system manipulated text directly and was given the initiative by the user. With \wws, only the user had the initiative.

\subsection{Generative Models can Inspire the User}

Our results show that both of our interaction methods helped to inspire the user (see \Cref{fig:questionnaire_inspiration} \& \Cref{subsec:results_usability_inspiration}). The results are almost similar, but users agreed on \wwss to be more inspiring. With \wws, we always presented three generated phrases as suggestions. Displaying multiple phrases is more inspirational than displaying automatically written text. This finding is in line with work by \cite{buschek21}.

Furthermore, people gave feedback on both interaction methods inspiring them. In order to inspire them, the text did not have to be an exact fit. It was mentioned oftentimes that even if the text did not fit, it gave impulses for new ideas on the topic or an improved wording.

This feedback shows that in order to be inspiring, generated text has not always to be perfect or be taken over as is. The offered text provides the user an idea for a new way of continuing the text. 
A similar result was found in the work of \citet{singh22}, who called people's (creative) integration of AI contributions ``integrative leaps''. They showed that various sources of inspiration help the user to keep writing when feeling stuck. In work by \citet{arnold21}, users kept track of the content of the text by giving answers on generated questions. This approach helped writers to make ideas more clear. These two works specifically researched generative AI with the purpose of inspiring the user. In comparison, we found inspiration as a side effect of writing with a generative model. Thus, we argue that inspiration occurs as a general effect when writing with generative AI. In this regard, we see the need for future studies, similar to \citet{singh22}, to explore how inspiring text AI should be integrated into UIs and how they might have to be concretely adapted to certain text types, e.g. formal vs. informal, or writing stages, e.g. drafting vs. iterating.

In general, inspiration is a subjective measure: For example, we asked people if a method was inspiring. It remains a challenge to find objective indicators for inspiration when writing with AI. For this purpose, analytical methods from NLP could be combined with qualitative text analysis to find such indicators. This is another research direction we see for future work on writing with inspiring AI.

\subsection{Initiative and Control Should be Integrated as Characteristics in Role Frameworks}

We have argued that the differences in initiative and control inherent in each design resulted in measurable changes in interaction. Furthermore, we see an interrelation between initiative and control, and roles when using the methods. We compare the inherent roles of our generative interaction methods to related work on co-creativity. In addition, we look at turn taking within the co-creative process. Before we go into detail for role comparison, we want to clarify that we did not implement an anthropomorphic UI to embody the generative model as an entity. We rather integrated the generative models into the UI as a tool. This clarification is important since some related work on co-creative roles frequently use the term agent -- in our case, the generative model (AI) may be exchangeable with the term agent. 
We rely on three frameworks to classify roles and to reflect on the process of co-creativity in our interaction methods. 

First, with the framework by \citet{negrete-yankelevich14}, we attribute the role of a \textit{generator} to both of our methods. Considering this related work, a \textit{generator} is defined by the capability to generate specimens or prototypes of partial or complete pieces of work. Further, we attribute the role of a \textit{master} to \cgts and the role of an \textit{apprentice} to \wws. These roles are underlined by how the methods are intended to work: In \cgts the system produces a text and the user rather manages and edits the generation, where in \wwss the system produces a set of suggestions and the user has to pick the best.

Second, the framework by \citet{kantosalo16} allows for a more fine-grained distinction between roles and capabilities. Regarding this framework, we would attribute the role of \textit{incomplete agents} to both methods, since they are capable of identifying a task and generating text, yet they do not have the capability of evaluating output. Further, we would attribute \cgts as taking part in \textit{alternating co-creativity}. The authors of the framework argue that an \textit{incomplete agent} cannot take part in alternative co-creativity. However, based on our experiences in this work, we think the missing criteria of a system to evaluate its own output will only cause a drop in the quality of the system's output, but the ability to participate in alternating co-creativity is preserved. For \wwss we attribute the role of \textit{task-divided co-creation}, since human and system do not take equal turns, instead the system only contributes suggestions. Finally, we use the framework to distinguish between the way an agent contributes to co-creativity. We consider \wwss as \textit{pleasing} since it conforms more to the user's intent, and \cgts as \textit{provoking} since it is challenging the user. The latter is also reflected in the ratings on helpfulness as seen in \Cref{fig:likert_ratings_helpfulnes}). \cgts was perceived as less helpful (more challenging) and \wwss as more helpful (more pleasing).

Third, with the framework by \citet{guzdial19}, we depict the process of co-creativity between human and the generative model: In the case of our prototypes, the users are people who are able to write text on smartphones. The AI is a generative model but static, it does not learn by taking into consideration previous interactions. The first turn of the interaction is taken by the user by starting writing the text. With \cgts the turn is then given to the AI by the user when the system generation is started. The system executes an \textit{artifact action} by writing text word by word, the user at the same time observes passively the system's output. The user can take the turn again if text deviates too much from expectation by stopping the generation. In an \textit{artifact action} the user can edit the text, the AI does not take any actions in the meanwhile. The user's turn ends again by starting the continuously generative writing again. In contrast, in \wws, the user always takes turn and executes \textit{artifact actions} by writing text. The AI executes only passive actions by suggestion completions of sentences. Selecting and taking over one of these suggestions is \textit{another artifact action} by the user. With both methods, the turns finally end if the user's expectations on the text outcome are met.

In all of these frameworks we find the dimensions of control and initiative implicitly again, be it in roles, e.g. master vs. apprentice, or interaction, e.g. alternating co-creativity, task-divided co-creation, or turn taking. Based on our experiences and this reflection, we propose to adopt \textit{initiative} and \textit{control} as explicit dimensions in any framework designed to determine roles. 
Overall, we see the need for a role and process framework for co-creative interaction with generative AI to inform future research and design directions.

\subsection{Perceived Quality and Wording Change with Text Written by Generative AI, Authorship is Preserved with Suggestions}

Using the generative interaction methods, people were less satisfied with the outcome of the text (see \Cref{fig:likert_ratings_satisfaction} \& \Cref{subsec:results_satisfaction_difficulty}). We interpret this rating on satisfaction as a drop in text quality. Furthermore, texts got longer when writing with AI (see \Cref{fig:result_textmetrics_textlength} \& \Cref{subsec:wpm_textlength}). We suspect this extended length grows proportional, the more text is written with generative AI. Furthermore, we suspect the extended length causes problems for formal text with a focus on information, since this additional text might cause a drop in information density overall. Our findings in this regard deviate from those by \citet{arnold20}. They reported that writing with suggestions makes text shorter. This might be true when writing image captions (their case), i.e. very short texts, with suggestions that are only one word long. However, suggestions can vary in length, for example, we employed suggestions with a length from eight to twelve words. We assume that integrating those partial sentences introduced more words than users would have intended in manual writing, and thus made text longer. This finding shows that we cannot generalize the statement by \citeauthor{arnold20} that writing with suggestions makes text longer. To refine the statement: The final text length depends on the length of adopted AI-generated text (e.g. suggestion length) and the text format. As a result of reduced satisfaction on text quality and the extended text length, we  recommend to revise a text written with generative AI at least once.

Overall, the texts did not only get longer, people also perceived a change in wording when writing with AI-generated text (see \Cref{fig:likert_ratings_wording} \& \Cref{subsec:results_authorship_wording}). However, when writing with \wws, people still perceived themselves as authors. This perception of authorship can be considered a contradiction, since people also had the least text input interactions with \wws. People realize that taking over generated suggestions changes wording, but do not realize it might also change ``authorship'' in terms of the length of human vs AI contributed text. We suspect that a high level of control over accepting generated text and having the initiative in writing are the reasons for this perception. We suggest future studies to investigate more fine-grained levels of authorship, for example, asking if users would attribute co-authorship to the system.

\subsection{Design and Research Implications}

Based on insights from our analysis on perceived authorship, we suggest raising awareness for authorship in generated text for users or use-cases where this is valued. Text that was generated by AI could be highlighted, for example, by changing font color or adding a background color behind the phrases. Visualizing the ratio of own and generated text would be another option in this regard. This information could be displayed next to other statistics, for example, next to the word count.

Furthermore, we recommend to make suggestions configurable, based on the ratings on suggestion length (see \Cref{fig:combined_ratings}). The users could then adjust the number of displayed suggestions, and also the length. In addition to that, suggestions should adapt to context. This context could be, for example, the writing stage. In an idea finding phase, the suggestions would be shorter and more diverse. After the idea finding phase, the suggestions could be longer and more informative. These adaptive suggestions would only be constrained by configurations. Moreover, such adaptions could be extended to the interaction design. The interaction methods could switch, for example, in an ideation phase \wwss could help with finding ideas and when ideas are found, \cgts could help to extend text. Another possibility could be to adapt the input methods to the roles, for example, when the user is actively writing, suggestions could be displayed, when the user gets more passive, the system could continue with writing. How to apply these configurations and adaptions, however, requires more research.

Considering our findings on text correction, we recommend to design control options for interacting with systems that automatically write text, e.g. to make the generation more directly steerable. For instance, a function to rewind the written text could be an option for the method \cgts to avoid long backspace sequences (see \Cref{fig:result_inputs_lengthbackspacesequences}). 

Our prototypes generated text with GPT-2. Similar to \citeauthor{singh22} we assume an improved quality of generated text by future models may influence how users interact with and perceive generative systems. This motivates further studies with other models.

In general, our interaction designs demonstrate diverging pathways of how text can be written with generative AI. Assuming further advances in the performance of language models, we expect that, on the one hand, generative text models will replace tedious routine writing (e.g. emails sent to agree on a meeting). On the other hand, we see the potential to aid human writing skills for more (co-)creative tasks (e.g. AI-supported novel writing). In a broad outlook, working on tasks together with generative AI will likely influence the future working life and productivity. 
\section{Conclusion}

We have examined writing with AI text generations on mobile devices: We compared two distinct methods in which users and AI can both contribute text when writing on mobile devices. Concretely, we built two prototypes that demonstrate 1) writing with AI generated suggestions and 2) a continuously writing AI. We provide the following findings as answers to our research question on how fundamental human-AI roles, as embedded in UI and interaction design, influence text entry behavior, experience, and output.

Based on our findings, we propose to take initiative as an explicit design dimension into consideration when designing and implementing human-AI interaction to result in truly mixed-initiative applications. Furthermore, we identified control and initiative as possible characteristics for frameworks that conceptualize human-AI roles and propose to extend such frameworks by these dimensions. Our analysis of text entry and behavior showed that people wrote less actively with AI suggestions, but still perceived themselves as authors. Instead, continuously written AI text reduced this feeling of authorship, while effort for editing increased. Overall, the AI generations increased text length and were perceived to influence wording. These results together yield our key conclusion and contribution to the literature: Users' perceived authorship of output, when co-created with AI, is not a simple function of the number of user actions. Instead, the types of these actions are crucially important (here: editing actions vs. typing actions), and those in turn result from the roles embedded in the interaction design.

Finally, we provide examples for practical design decisions and discuss broader research implications. If model performance improves, we assume that generative models might replace tedious routine writing in the future. At the same time, they will aid human writing skills for creative and original text. In this light, our study highlights the importance to better understand the implications of designing human-AI interaction for writing creative texts with generative models. The overarching goal of our research is thus to contribute to future human-AI interaction that enhances human skill, instead of replacing it.

\begin{acks}
This project is funded by the Bavarian State Ministry of Science and the Arts and coordinated by the Bavarian Research Institute for Digital Transformation (bidt).
\end{acks}

\bibliographystyle{ACM-Reference-Format}
\bibliography{literature}

\onecolumn

\section{Appendix}

\subsection{Likert Questions} \label{subsec:appendix-likert}

\begin{table}[htb]
\centering
    \begin{tabularx}{\textwidth}{m{0.03\textwidth} | m{0.55\textwidth} | m{0.35\textwidth}}
            \textbf{ID} & \textbf{Question}                                 & \textbf{Response type} \\  \hline \hline
            1  & I am satisfied with the text. & \multirow{6}{0.35\textwidth}{5-point-scale\\ \small{(1=Strongly disagree, 5=Strongly agree)}} \\ \cline{1-2}
            2  & It was easy for me to write the text.                      & \\ \cline{1-2}
            3  & I feel like I am the author of the text.                   & \\ \cline{1-2}
            4  & The interaction method was suitable for this task.         & \\ \cline{1-2}
            5  & The interaction method helped me write the text.           & \\ \cline{1-2}
            6  & The interaction method influenced the wording of the text. & \\
    \end{tabularx}
    \caption{Likert questions, we asked people after completing a task.}
    \label{tab:likert-questions}
\end{table}

\subsection{Questionnaire} \label{subsec:appendix-questionnaire}

\begin{table}[htb]
\centering
    \begin{tabularx}{\textwidth}{p{0.03\textwidth} | p{0.55\textwidth} | p{0.35\textwidth}}
            \textbf{ID} & \textbf{Question} & \textbf{Response type} \\  \hline \hline
            A1  & Enter your given user id.                            
                & Free-text \\ \hline
            A2  & What is your gender?                     
                & \makecell[tl]{Selection \\
                    \small{(Woman, Man, Non-binary,} \\ 
                    \small{Prefer not to disclose,}  \\ 
                    \small{Prefer to self-describe)}} \\ \hline
            A3  & What is your Age?                   
                & Number \\ \hline
            A4  & What is your English level (based on the CEFR Scale)?         
                & \makecell[tl]{Selection \\
                    \small{(C2, C1, B2, B1, A2, A1)}} \\ \hline
            A5  & Which is your dominant hand?         
                & \makecell[tl]{Selection \\ 
                    \small{(Right, Left, No dominant hand)}} \\ \hline
            A6  & How do you typically type on your mobile phone? 
                & \makecell[tl]{Selection \\ 
                    \small{(One handed with the thumb,} \\ 
                    \small{One handed with the index finger,} \\ 
                    \small{Two handed with both thumbs, Other)}} \\ \hline
            A7  & What kind of mobile phone did you use for the study? 
                & Free-text \\ \hline
            A8  & Did you have any previous experience with writing with generated text? (Using suggestions etc. on mobile devices also counts.)
                & \makecell[tl]{Selection \\ 
                    \small{(Yes, No)}} \\ \hline
            A9  & If yes, list all previous experiences which you had with writing with generated text. 
                & Free-text \\
    \end{tabularx}
    \caption{Demographics}
    \label{tab:demographics}
\end{table}

\begin{table}[htb]
\centering
    \begin{tabularx}{\textwidth}{p{0.03\textwidth} | p{0.55\textwidth} | p{0.35\textwidth}}
            \textbf{ID} & \textbf{Question} & \textbf{Response type} \\  \hline \hline
            B1  & Which of the two interaction methods, working with generated text, did you prefer for the task “It's your friend's birthday. Write an e-mail to wish all the best and mention that you have to meet again in some time.“?                
                & \makecell[tl]{Selection \\
                    \small{(Continuous generated text,} \\ 
                    \small{Writing with suggestions)}}  \\  \hline
                    
            B2  & And shortly explain why you preferred this method for this task.                     
                & Free-text \\ \hline
                
            B3  & For this task (“It's your friend's birthday. Write an e-mail to wish all the best and mention that you have to meet again in some time.“), would you prefer the standard input method over the method with generated text you chose before? 
                & \makecell[tl]{Selection \\ 
                    \small{(Yes, No)}} \\ \hline
                    
            B4  & And can you shortly explain your decision?       
                & Free-text \\ \hline
                
            B5  & Which of the two interaction methods, working with generated text, did you prefer for the task “Write a short story about your last or upcoming vacation.“?       
                & \makecell[tl]{Selection \\
                    \small{(Continuous generated text,} \\ 
                    \small{Writing with suggestions)}}  \\  \hline
                    
            B6  & And shortly explain why you preferred this method for this task.                     
                & Free-text \\ \hline
                
            B7  & For this task (“Write a short story about your last or upcoming vacation.”), would you prefer the standard input method over the method with generated text you chose before?
                & \makecell[tl]{Selection \\ 
                    \small{(Yes, No)}} \\ \hline
                
            B8  & And can you shortly explain your decision?       
                & Free-text \\
    \end{tabularx}
    \caption{Questions about the tasks}
    \label{tab:questions-tasks}
\end{table}

\begin{table}[htb]
\centering
    \begin{tabularx}{\textwidth}{p{0.03\textwidth} | p{0.55\textwidth} | p{0.35\textwidth}}
            \textbf{ID} & \textbf{Question} & \textbf{Response type} \\  \hline \hline
            C1  & How hard was it to learn and use this method?               
                & \makecell[tl]{5-point-scale \\
                    \small{(1=Very easy, 5=Very hard)}} \\  \hline
                    
            C2  & How was the speed of the appearing text?                    
                & \makecell[tl]{5-point-scale \\
                    \small{(1=Way too slow, 5=Way too fast)}} \\  \hline
                
            C3  & The interaction method inspired me to write things I normally wouldn't have thought of.
                & \makecell[tl]{5-point-scale \\ 
                    \small{(1=Strongly disagree, 5=Strongly agree)}} \\ \hline
                    
            C4  & I think the inspiration for new formulations / different wordings, provided by this interaction method, could be helpful in some cases. (Assuming the model would be better)      
                & \makecell[tl]{5-point-scale \\ 
                    \small{(1=Strongly disagree, 5=Strongly agree)}} \\ \hline
                
            C5  & In which situation/use-case could this interaction method be useful?      
                & Free-Text \\  \hline
                    
            C6  & What are the advantages of this method relating to the writing interaction?                    
                & Free-text \\ \hline
                
            C7  & What are the disadvantages of this method relating to the writing interaction?
                & Free-text \\ \hline
                
            C8  & Can you imagine any features which could make this method better?     
                & Free-text \\
    \end{tabularx}
    \caption{Questions about “Continuous generated text”}
    \label{tab:questions-cgt}
\end{table}

\begin{table}[htb]
\centering
    \begin{tabularx}{\textwidth}{p{0.03\textwidth} | p{0.55\textwidth} | p{0.35\textwidth}}
            \textbf{ID} & \textbf{Question} & \textbf{Response type} \\  \hline \hline
            D1  & How hard was it to learn and use this method?               
                & \makecell[tl]{5-point-scale \\
                    \small{(1=Very easy, 5=Very hard)}} \\  \hline
                    
            D2  & How was the number of suggestions provided?                  
                & \makecell[tl]{5-point-scale \\
                    \small{(1=Way too low, 5=Way too high)}} \\  \hline
              
            D3  & How was the length of each suggestion?                 
                & \makecell[tl]{5-point-scale \\
                    \small{(1=Way too short, 5=Way too long)}} \\  \hline   
                
            D4  & The interaction method inspired me to write things I normally wouldn't have thought of.
                & \makecell[tl]{5-point-scale \\ 
                    \small{(1=Strongly disagree, 5=Strongly agree)}} \\ \hline
                    
            D5  & I think the inspiration for new formulations / different wordings, provided by this interaction method, could be helpful in some cases. (Assuming the model would be better)      
                & \makecell[tl]{5-point-scale \\ 
                    \small{(1=Strongly disagree, 5=Strongly agree)}} \\ \hline
                
            D6  & In which situation/use-case could this interaction method be useful?      
                & Free-Text \\  \hline
                    
            D7  & What are the advantages of this method relating to the writing interaction?                    
                & Free-text \\ \hline
                
            D8  & What are the disadvantages of this method relating to the writing interaction?
                & Free-text \\ \hline
                
            D9  & Can you imagine any features which could make this method better?     
                & Free-text \\
    \end{tabularx}
    \caption{Questions about “Writing with suggestions”}
    \label{tab:questions-wws}
\end{table}

\end{document}